\documentclass[journal=jpcbfk,manuscript=article]{achemso}
\setkeys{acs}{articletitle = true}
\setkeys{acs}{chaptertitle = true}
\setkeys{acs}{maxauthors = 0}
\setkeys{acs}{keywords = true}
\usepackage{chemformula} 
\usepackage[T1]{fontenc} 
\usepackage{braket}
\usepackage{amsmath,amssymb}
\usepackage{ulem}        

\author{Wataru Ota}
\affiliation{Fukui Institute for Fundamental Chemistry, Kyoto University, Sakyo-ku, Kyoto 606-8103, Japan}
\alsoaffiliation{Department of Molecular Engineering, Graduate School of Engineering, Kyoto University, Nishikyo-ku, Kyoto 615-8510, Japan}

\author{Ken Takahashi}
\affiliation{Undergraduate School of Industrial Chemistry, Faculty of Engineering, Kyoto University, Nishikyo-ku, Kyoto 615 8510, Japan}

\author{Kenji Higashiguchi}
\affiliation{Department of Synthetic Chemistry and Biological Chemistry, Kyoto University, Nishikyo-ku, Kyoto 615-8510, Japan}

\author{Kenji Matsuda}
\affiliation{Department of Synthetic Chemistry and Biological Chemistry, Kyoto University, Nishikyo-ku, Kyoto 615-8510, Japan}

\author{Tohru Sato}
\affiliation{Fukui Institute for Fundamental Chemistry, Kyoto University, Sakyo-ku, Kyoto 606-8103, Japan}
\alsoaffiliation{Department of Molecular Engineering, Graduate School of Engineering, Kyoto University, Nishikyo-ku, Kyoto 615-8510, Japan}
\alsoaffiliation{Unit of Elements Strategy Initiative for Catalysts \& Batteries, Kyoto University, Nishikyo-ku, Kyoto 615-8510, Japan}
\email{tsato@scl.kyoto-u.ac.jp}

\title
{
Origin of Aggregation-Induced Enhanced Emission:\\
Role of Pseudo-Degenerate Electronic States of Excimers Formed in Aggregation Phases
}

\abbreviations{AIEE,OLED,HPS,TADF,RISC,CNBE,VC,VCC,VCD,HOMO,LUMO,TD-DFT,ONIOM,QM,MM,PCM}

\begin{document}


\begin{abstract}
Origin of aggregation-induced enhanced emission (AIEE) is
investigated considering cyano-substituted 1,2-bis(pyridylphenyl)ethene (CNPPE) as an example.
On the basis of ONIOM calculations using the time-dependent density functional theory (TD-DFT),
we found that pseudo-degeneracy of excimers formed in solid phase plays an important role
in the appearance of AIEE.
The electron density difference delocalized over molecules
gives rise to small diagonal vibronic coupling constants (VCCs),
which suggests that the internal conversion is more suppressed in solid phase than in solution phase.
The reduction of the off-diagonal VCCs owing to the packing effect
is elucidated by vibronic coupling density (VCD) analysis.
The pseudo-degeneracy enables fluorescence from the high singlet excited states 
against Kasha's rule
because the electron density difference and the overlap density 
between the excited states vanish.
A Hubbard model of a pseudo-degenerate electronic system is constructed to
explain the vanishing mechanism.
We propose the following design principle for AIEE:
\textit{
a candidate molecule for AIEE should have pseudo-degenerate adiabatic electronic states 
in the aggregation phases originating from the excimer formation.
}
\end{abstract}


\section{Introduction}
In contrast to concentration quenching,
aggregation-induced enhanced emission (AIEE) 
yields strong luminescence in the aggregation phases
\cite{Hong2009_4332,Hong2011_5361}.
The restriction of intramolecular motions is generally accepted as the reason behind AIEE,
that is, the restriction of intramolecular rotations or vibrations 
arising from the physical constraint in the aggregation phases
blocks non-radiative transition pathways
\cite{Mei2014_5429}.
Hexaphenylsilole (HPS) is one of the AIEE dyes, 
and the restricted rotation of the side phenyl ring is found to be a key factor for AIEE
\cite{Chen2003_1535,Yu2005_6335,Zhao2015_5347}.
This is corroborated by quantum mechanical and molecular mechanical (QM/MM) calculations 
where the Huang--Rhys factors and reorganization energies at low-frequency modes 
are reduced in the solid phase compared to the gas phase owing to the packing effect
\cite{Zhang2014_9094}.
In addition, HPS having a bulky shape does not form a cofacial configuration in the solid phase
\cite{Hong2009_4332,Hong2011_5361},
which is considered to prevent concentration quenching.

Cofacial aggregation is regarded as one of the reasons behind concentration quenching.
This is because a destructive alignment of the transition dipole moment
renders the lowest excited state that is symmetry forbidden
\cite{Cornil2001_1053}.
Nevertheless, some dyes have been reported to be emissive 
in spite of the formation of cofacial aggregations
\cite{Rosch2006_7184,Yoon2010_13675,Yao2011_834,Wang2014_8723,
Basak2015_30398,Lucenti2017_1894,Qian2017_83,Ryu2017_8870}.
In general, 
according to Kasha's rule,
emissions have been attributed to the lowest excited states
\cite{Kasha1950_14}.
However, 
if all the radiative and non-radiative transitions 
from a higher excited state to all the lower excited states are suppressed,
fluorescence from the higher excited state can be expected to occur
against Kasha's rule.

Sato {\it et al.} have already reported that, 
in fluorescent dopants employed in organic light-emitting diodes (OLEDs),
radiative and non-radiative transitions from a triplet excited state T$_n$ ($n>1$) 
to all the lower triplet excited states can be suppressed 
due to the pseudo-degenerate electronic states
\cite{Sato2015_189,Sato2017_4820,Pu2019_2541}.
The pseudo-degeneracy leads to cancellation of the overlap density between the excited states,
which generates T$_n$ excitons with long lifetimes.
This enables the fluorescence via higher triplets (FvHT) mechanism for OLEDs,
that is fluorescence utilizing the reverse intersystem crossing (RISC) 
from T$_n$ to singlets.
FvHT is different from thermally activated delayed fluorescence (TADF)
in that TADF undergoes the thermally activated RISC from T$_1$ to S$_1$ 
by decreasing the energy difference, $\Delta E_{{\rm S_1-T_1}}$
\cite{Endo2009_4802,Adachi2014_060101}.
The FvHT mechanism is proposed to explain the mechanism of electroluminescence
in some OLED dopants with large $\Delta E_{{\rm S_1-T_1}}$,
where T$_1$ excitons cannot overcome the energy difference thermally
\cite{Hu2014_2064,Sato2017_4820}.

\begin{figure}[h]
\centering
    \includegraphics[width=0.8\hsize]{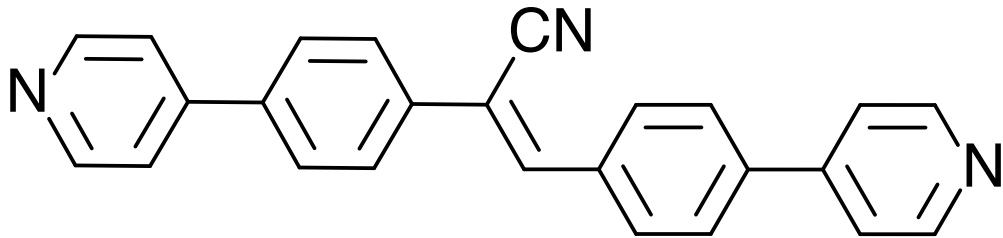}
  \caption{
	Chemical structure of cyano-substituted 1,2-bis(pyridylphenyl)ethene (CNPPE).
	   }
  \label{FIG1}
\end{figure}

A cyano-substituted 1,2-bis(pyridylphenyl)ethene (CNPPE) (Fig.~\ref{FIG1})
has been reported to exhibit the AIEE behavior in solid phase
\cite{Nishio2014_686}. 
The rate constants of the non-radiative transitions are decreased
from $> 1.0\times10^{10}$ s$^{-1}$ in CH$_2$Cl$_2$ solution
to $5.0\times10^7$ s$^{-1}$ in solid phase.
The rate constants of the radiative transitions are comparable 
between CH$_2$Cl$_2$ solution ($> 2.0\times10^7$ s$^{-1}$) and
solid phase ($1.3\times10^8$ s$^{-1}$).
Accordingly, the fluorescence quantum yield is increased from 0.002 to 0.72 by aggregation.
Since CNPPE forms cofacial configurations in solid phase,
the occurrence of concentration quenching is predicted.
Some cofacial CNPPE molecules have $C_i$ symmetry in the crystal structure.
This suggests the possibility of pseudo-degenerate electronic states 
delocalized over the cofacial molecules.
In this case, 
fluorescence from higher singlets than S$_1$ is expected to occur against Kasha's rule
because the transitions between the excited states can be suppressed
as in the case of the FvHT mechanism.
The electronic states delocalized over molecules
may decrease the vibronic coupling constants (VCCs) 
\cite{Shizu2013_215}.
These indicate that the internal conversion can be more suppressed in solid phase
than in solution phase as long as excimer formation occurs in solid phase.
This mechanism is different from what is proposed in styrylbenzene, i.e.
di
\cite{Shi2017_23166}, 
tri
\cite{Garzon2017_4720,Domiguez2020_1},
and tetra(styryl)benzene
\cite{Domiguez2020_1},
which asserts that the origin of the AIEE behavior is attributed to 
blocking the trans-cis photoisomerization of stilbene unit in solid phase.

In this study, we investigated the role of pseudo-degeneracy in the appearance of AIEE
considering CNPPE as an example.
Vibronic coupling density (VCD) analyses were performed to elucidate the local picture of VCC
\cite{Sato2008_758,Sato2009_99,Sato2013_012010}.
To explain the results obtained by the TD-DFT calculations, 
we discuss the electron density difference and overlap density 
in the pseudo-degenerate electronic system based on the Hubbard model.

\section{Theory}
We consider the internal conversion from an initial vibronic state 
$\ket{\Phi_{mi} ({\bf r}, {\bf Q})}$ 
associated with electronic $m$ and vibrational $mi$ states
to a final vibronic state $\ket{\Phi_{nj} ({\bf r}, {\bf Q})}$.
Here, ${\bf r}=({\bf r}_1, \cdots, {\bf r}_i, \cdots {\bf r}_N)$ 
is a set of $N$ electronic coordinates,
and ${\bf Q} = (Q_1, \cdots, Q_{\alpha}, \cdots Q_{M})$ 
is a set of $M$ mass-weighted normal coordinates.
Within the crude adiabatic approximation
\cite{Fischer1984,Azumi1977_315},
the vibronic states are represented as
the product of vibrational and electronic states fixed at the nuclear configuration ${\bf R}_0$:
$\ket{\Phi_{mi} ({\bf r}, {\bf Q})} = \ket{\chi_{mi} ({\bf Q})} \ket{\Psi_m ({\bf r}; {\bf R}_0)}$.
${\bf R}_0$ is chosen as the equilibrium nuclear configuration 
of the ground or excited optimized structures.
The rate constant of the internal conversion 
from electronic state $m$ to $n$ is expressed as
\cite{Uejima2014_14244}
\begin{equation}
	k_{n \leftarrow m}^{{\rm IC}} (T) 
	= \frac{2\pi}{\hbar} \sum_{\alpha} |V_{mn,\alpha}|^2
	\sum_{ij} P_{mi} (T) |\braket{\chi_{mi}|Q_{\alpha}|\chi_{nj}}|^2
	\delta (E_{mi} - E_{nj}),
\end{equation}
where 
$P_{mi}(T)$ is the Boltzmann distribution function of the initial vibronic state at temperature $T$, 
and $E_{mi}$ and $E_{nj}$ are the eigenenergies of 
$\ket{\Phi_{mi} ({\bf r}, {\bf Q})}$ and $\ket{\Phi_{nj} ({\bf r}, {\bf Q})}$, respectively.
$E_{mi}$ is represented as the sum of electronic $E_m$ and vibrational energies.
$V_{mn, \alpha}$ is the off-diagonal VCC given by
\begin{equation}
	V_{mn, \alpha} = 
	\left< \Psi_m ({\bf r}; {\bf R}_0) \left| 
	\left( \frac{\partial \hat{H} ({\bf r}, {\bf R}) }{\partial Q_{\alpha}} \right)_{{\bf R}_0} \right|
	\Psi_n ({\bf r}; {\bf R}_0) \right>,
\end{equation}
where $\hat{H} ({\bf r}, {\bf R})$ is the molecular Hamiltonian, 
and ${\bf R}$ is a set of nuclear configuration.
$V_{n}:=V_{nn}$ is called the diagonal VCC.
Ignoring the Duschinsky effect,
which means vibrational modes do not change during an excitation,
the matrix element of the vibrational states is written as 
\begin{equation}
	\braket{\chi_{mi}|Q_{\alpha}|\chi_{nj}}
	= \braket{n_{mi,\alpha}|Q_{\alpha}|n_{nj, \alpha}} 
	\prod_{\beta \neq \alpha} \braket{n_{mi,\beta}|n_{nj,\beta}},
\end{equation}
where $\ket{n_{mi,\alpha}}$ is a vibrational state of a single mode.
The Franck--Condon (FC) overlap integral is expressed as
\cite{Hutchisson1930_410}
\begin{eqnarray}
	& & \braket{n_{mi,\alpha}|n_{nj,\alpha}}
	= \sqrt{\frac{ n_{mi, \alpha}! n_{nj, \beta}!}{2^{n_{mi,\alpha}+n_{nj,\alpha}}}}
	e^{- \frac{1}{4} g_{n,\alpha}^2} \notag \\
	& \times & \sum_{l=0}^{ {\rm min} [n_{mi,\alpha} n_{nj,\alpha}]}
	(-1)^{n_{mi,\alpha}-l} 2^l 
	\frac{ g_{n,\alpha}^{n_{mi,\alpha}+n_{nj,\alpha}-2l} }{ l! (n_{mi,\alpha}-l)! (n_{nj,\alpha}-l)!}.
\end{eqnarray}
Here, $g_{n,\alpha}$ is the dimensionless diagonal VCC (the Huang--Rhys factor):
\begin{equation}
	g_{n,\alpha} = \frac{ V_{n,\alpha} }{ \sqrt{\hbar \omega_{n,\alpha}^3}},
\end{equation}
where $\omega_{n,\alpha}$ is the angular frequency of vibrational mode $\alpha$.
In general, a rate constant of an internal conversion is small 
when diagonal and off-diagonal VCCs are small
\cite{Uejima2014_14244}.
Particularly, the dependence of the rate constant on the diagonal VCC is strong,
and the reduction of the diagonal VCCs significantly contributes to 
the suppression of the internal conversion
\cite{Uejima2014_14244}.

Within a single mode approximation at 0 K,
the rate constant of the internal conversion is reduced to
\cite{Uejima2014_14244}
\begin{equation}
	k_{n \leftarrow m}^{{\rm IC}} (T) 
	= \frac{2\pi}{\hbar} |V_{mn,\alpha}|^2
	\sum_j |\braket{\chi_{mi}|Q_{\alpha}|\chi_{nj}}|^2
        \int_{E_{{\rm min}}}^{E_{{\rm max}}} 
        \frac{1}{\sqrt{2 \pi} \sigma} e^{ -\frac{(E - (E_{mi}-E_{nj}))^2}{2\sigma^2} }
	dE,
  	\label{Eq:IC2}
\end{equation}
where the delta function is replaced with the Gaussian function 
which represents the density of states with the linewidth $\sigma$.

In addition to the internal conversion,
vibrational relaxation from FC to adiabatic (AD) states is a non-radiative process
that should suppress for emission
\cite{Uejima2014_14244}.
Within the crude adiabatic representation assuming the harmonic approximation,
the reorganization energy due to vibrational relaxation is evaluated by
\begin{equation}
	\Delta E = \sum_{\alpha} \frac{V_{n,\alpha}^2}{2 \omega_{n,\alpha}^2}.
	\label{Eq:DeltaE}
\end{equation}
Thus, the reduction of the diagonal VCCs leads to the small reorganization energy
that depends on the square of the diagonal VCCs.

The VCD is expressed as the density form of the VCC
\cite{Sato2008_758,Sato2009_99,Sato2013_012010}:
\begin{equation}
	V_{mn, \alpha} 
	= \int d \textbf{\textit{x}} \ \eta_{mn,\alpha} (\textbf{\textit{x}}),
\end{equation}
where $\textbf{\textit{x}}=(x,y,z)$ is a three dimensional coordinate.
The diagonal VCD 
$\eta_{n, \alpha} (\textbf{\textit{x}}) := \eta_{nn,\alpha} (\textbf{\textit{x}})$ is defined by 
\begin{equation}
	\eta_{n,\alpha} (\textbf{\textit{x}}) 
	= \Delta \rho_{nm} (\textbf{\textit{x}}) \times v_{\alpha} (\textbf{\textit{x}}).
\end{equation}
Here, $\Delta \rho_{nm} (\textbf{\textit{x}})$ is the electron density difference 
between $\ket{\Psi_n ({\bf r}; {\bf R}_0)}$ and 
the reference state $\ket{\Psi_m ({\bf r}; {\bf R}_0)}$:
\begin{equation}
        \Delta \rho_{nm} (\textbf{\textit{x}}) = 
        \braket{\Psi_n ({\bf r};{\bf R}_0) | 
	\hat{\rho}(\textbf{\textit{x}}) | \Psi_n ({\bf r}; {\bf R}_0)} -
        \braket{\Psi_m ({\bf r};{\bf R}_0) | \hat{\rho}(\textbf{\textit{x}}) | 
        \Psi_m ({\bf r}; {\bf R}_0)}.
\end{equation}
$\ket{\Psi_m ({\bf r}; {\bf R}_0)}$ is taken as the electronic state 
in the equilibrium nuclear configuration.
$\hat{\rho}(\textbf{\textit{x}})$ is the electron density operator defined by
\begin{equation}
	\hat{\rho} (\textbf{\textit{x}}) 
	= \sum_{ij} \sum_{\sigma \tau} \hat{c}_{i \sigma}^{\dagger} \hat{c}_{j \tau}
	\psi_{i \sigma}^{*} (\textbf{\textit{x}}) \psi_{j \tau} (\textbf{\textit{x}}),
\end{equation}
where $\psi_{i \sigma}^{*} (\textbf{\textit{x}})$ and $\psi_{j \tau} (\textbf{\textit{x}})$ 
are spatial orbitals,
and $\hat{c}_{i \sigma}^{\dagger}$ and $\hat{c}_{j \tau}$ 
are creation and annihilation operators, respectively.
$i$ and $j$ refer to the orbital indices,
and $\sigma$ and $\tau$ refer to the spin indices.
$v_{\alpha} (\textbf{\textit{x}})$ is the potential derivative expressed as
\begin{equation}
	 v_{\alpha} (\textbf{\textit{x}}) 
	= \left( \frac{\partial u (\textbf{\textit{x}})}{\partial Q_{\alpha}} \right)_{{\bf R}_0},
\end{equation}
where $u(\textbf{\textit{x}})$ is the electron-nucleus potential acting on a single electron.

The off-diagonal VCD is defined by 
\begin{equation}
	\eta_{mn,\alpha} (\textbf{\textit{x}}) 
	= \rho_{mn} (\textbf{\textit{x}}) \times v_{\alpha} (\textbf{\textit{x}}),
\end{equation}
where $\rho_{mn} (\textbf{\textit{x}})$ is the overlap density 
between $\ket{\Psi_m({\bf r}; {\bf R}_0)}$ and $\ket{\Psi_n({\bf r}; {\bf R}_0)}$:
\begin{equation}
        \rho_{mn} (\textbf{\textit{x}}) 
        = \braket{\Psi_m({\bf r}; {\bf R}_0) | \hat{\rho} (\textbf{\textit{x}}) | \Psi_n({\bf r}; {\bf R}_0)}.
\end{equation}
The VCD enables us to understand the vibronic couplings (VCs)
arising from the electronic factor
$\Delta \rho_{nm} (\textbf{\textit{x}})$ or $\rho_{mn} (\textbf{\textit{x}})$ 
and the vibrational factor $v_{\alpha} (\textbf{\textit{x}})$.
It should be noted that
the disappearance of $\Delta \rho_{nm} (\textbf{\textit{x}})$ and $\rho_{mn} (\textbf{\textit{x}})$
gives rise to the suppression of the internal conversions 
via the reduction of $V_{n,\alpha}$ and $V_{mn,\alpha}$,
respectively.
Since the transition dipole moment also depends on $\rho_{mn} (\textbf{\textit{x}})$
\cite{Uejima2014_14244},
the disappearance of $\rho_{mn} (\textbf{\textit{x}})$ suppresses 
both the radiative and non-radiative transitions.

\section{Methods of calculations}

\begin{figure}[h]
\centering
    \includegraphics[width=0.6\hsize]{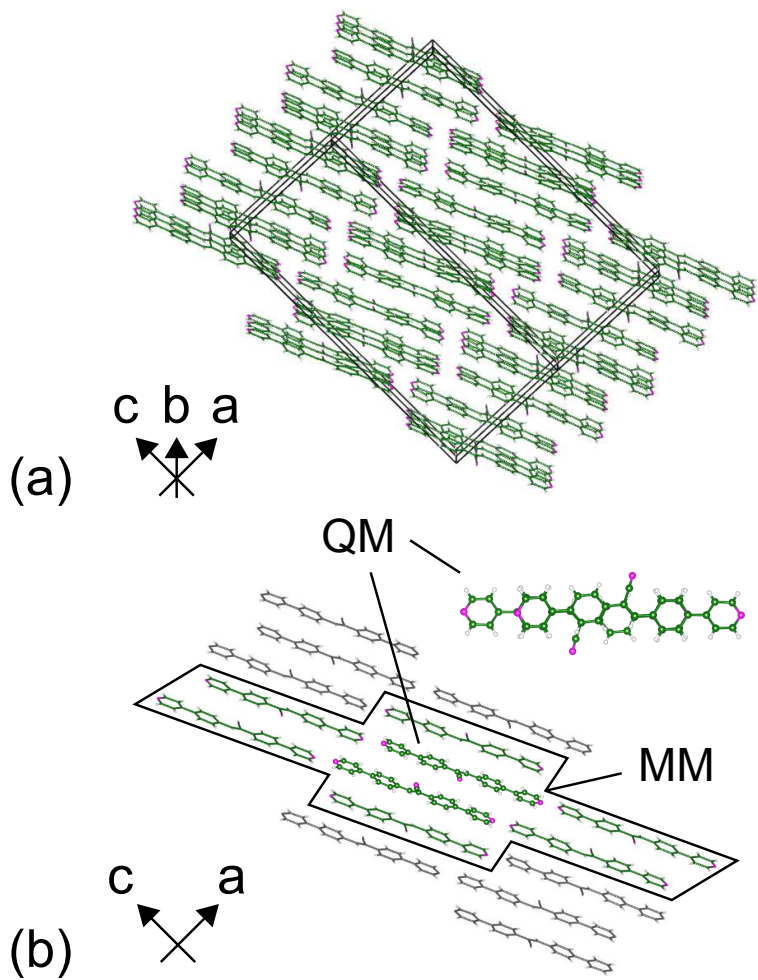}
  \caption{
   (a) Crystal structure of CNPPE solid
   \cite{Nishio2014_686}. 
   (b)
   Dimer Model \textbf{1} for the CNPPE solid where
   centered cofacial molecules with $C_i$ symmetry are selected as the QM region
   and the surrounding 16 molecules are selected as the MM region.
	   }
  \label{FIG2}
\end{figure}

Figure~\ref{FIG2} (a) shows the crystal structure of the CNPPE solid
\cite{Nishio2014_686}.
We modeled the CNPPE solid as a dimer with a cofacial configuration (Fig.~\ref{FIG2} (b)),
Dimer Model \textbf{1},
where the cofacial dimer was calculated by the QM method
and the surrounding 16 molecules were calculated by the MM method
based on the ONIOM 
(our own $n$-layered integrated molecular orbital and molecular mechanics)
approach
\cite{Svensson1996_19357,Chung2015_5678}.
Dimer Model \textbf{1} has $C_i$ symmetry.
The other dimer models aligned in different directions, 
Dimer Model \textbf{2} and Dimer Model \textbf{3} 
(see Fig. S1 in the Supplementary Information), 
having $C_1$ symmetry were also investigated.
The ground and excited states of the QM region were computed
at the M06-2X/6-31G(d,p) and TD-M06-2X/6-31G(d,p) levels of theory, respectively, 
whereas the MM region was computed using the universal force field (UFF).
The M06-2X functional was employed to incorporate dispersion interactions between adjacent molecules
\cite{Zhao2008_215}.
The coordinates of the MM region were fixed during the geometry optimizations and vibrational analyses
for the ground and excited states.

The CNPPE in CH$_2$Cl$_2$ solution was modeled as a single molecule with $C_1$ symmetry.
The ground and excited states of a single molecule in solution, Monomer Model, 
were computed at the M06-2X/6-31G(d,p) and TD-M06-2X/6-31G(d,p) levels of theory, respectively,
including the solvent effect through the polarizable continuum model (PCM)
\cite{Tomasi2005_2999}.
The above calculations were carried out using the Gaussian 09
\cite{Frisch2013D,Frisch2013E}.
The VCCs and VCD were calculated using our own code.

\section{Results and discussion}

\subsection{Vibronic coupling constants (VCCs) and vibronic coupling density (VCD)}

There are a few possibilities for selecting dimers from the crystal structure.
The total energies of excited states of the three types of dimer models were compared (Fig. S2).
The S$_1$ and S$_2$ states of Dimer Model \textbf{1} 
are energetically more stable than those of the other dimer models,
and the fluorescence from Dimer Model \textbf{1} is expected.
Therefore, we concentrate ourselves on Dimer Model \textbf{1}.
Table~\ref{TABLE1} lists the excited states of Dimer Model \textbf{1}
at the S$_0$ and S$_2$ optimized structures.
From the selection rule of the electric dipole transition, 
S$_1$ ($A_g$) is symmetry-forbidden
and S$_2$ ($A_u$) is symmetry-allowed (Laport\'{e} rule).
Although, according to Kasha's rule, 
an emission does not occur from the second excited states,
the fluorescence from S$_2$ is possible 
if all the transitions from S$_2$ to S$_1$ are suppressed.

\begin{table}[h]
\small
\caption{\label{TABLE1} 
Excited states of Dimer Model \textbf{1} at the S$_0$ and S$_2$ optimized structures.
$f$ denotes the oscillator strength.
}
  \begin{tabular*}{0.48\textwidth}{@{\extracolsep{\fill}}cccccc}
    \hline
         &               & \multicolumn{2}{c}{Excitation Energy} &         & Major Configuration \\
 \cline{3-4} 
         & State         & eV     &  nm                          & $f$     & (CI coefficient)\\
 \hline 
      @S$_0$ & S$_1$(A$_g$)  & 3.7359 & 331.88                  &  0.0000 & HO-1 $\rightarrow$ LU+1 (0.306)\\
             &               &        &                         &         & HO $\rightarrow$ LU (0.622) \\
             & S$_2$(A$_u$)  & 3.9052 & 317.49                  &  2.2807 & HO-1 $\rightarrow$ LU (0.366)\\
             &               &        &                         &         & HO $\rightarrow$ LU+1 (0.585) \\
      @S$_2$ & S$_1$(A$_g$)  & 3.1397 & 394.89                  &  0.0000 & HO-1 $\rightarrow$ LU+1 (0.212)\\
             &               &        &                         &         & HO $\rightarrow$ LU (0.664) \\
             & S$_2$(A$_u$)  & 3.3074 & 374.86                  &  1.8099 & HO-1 $\rightarrow$ LU (0.215)\\
             &               &        &                         &         &    HO $\rightarrow$ LU+1 (0.663)\\
\hline
  \end{tabular*}
\end{table}

\begin{figure}[h]
\centering
    \includegraphics[width=0.8\hsize]{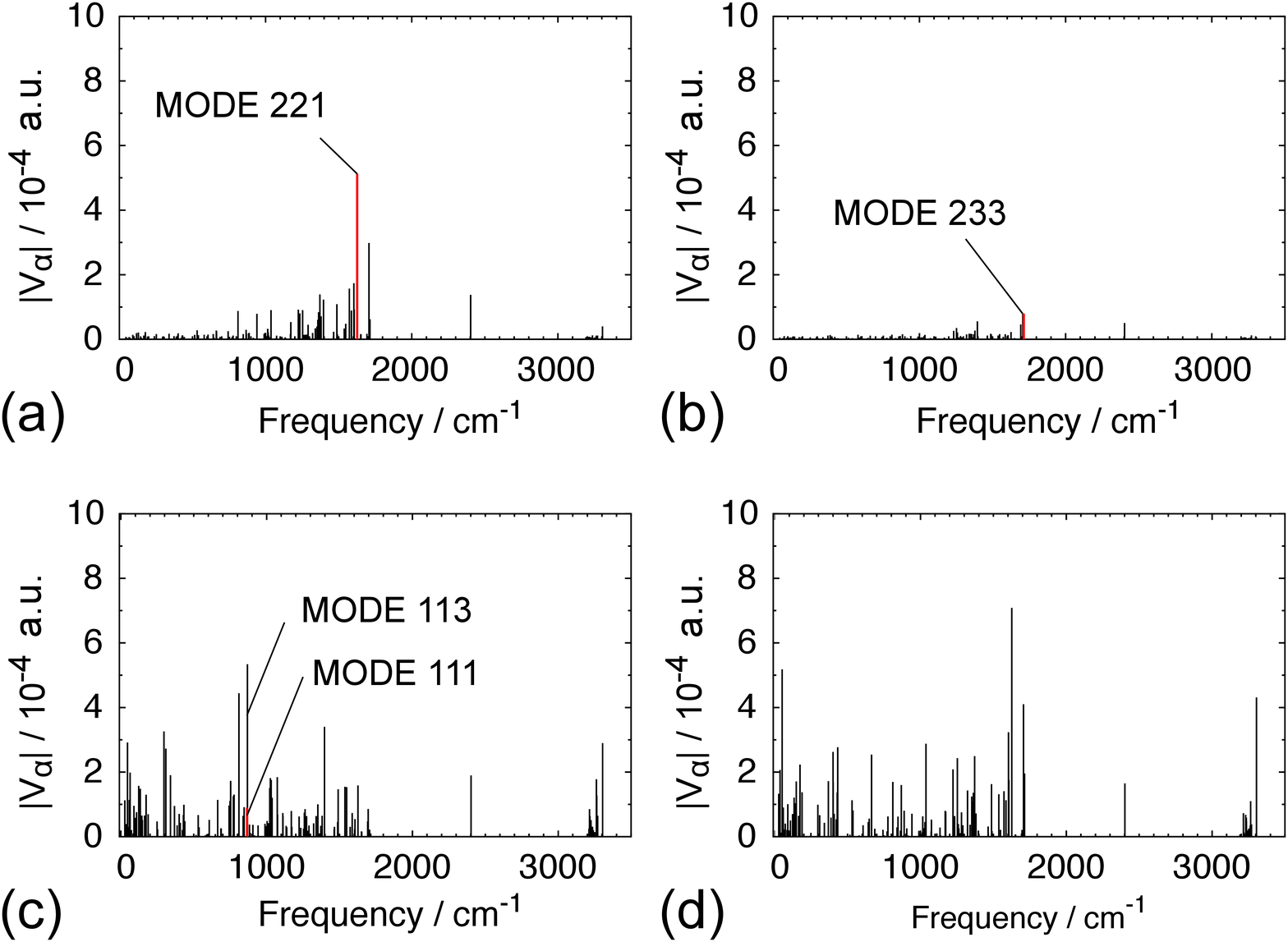}
  \caption{
	Diagonal VCCs of (a) S$_0$@S$_2$ and (b) S$_1$@S$_2$ as well as
	off-diagonal VCCs of 
	(c) S$_0$@S$_2$ $\leftarrow$ S$_2$@S$_2$ and (d) S$_1$@S$_2$ $\leftarrow$ S$_2$@S$_2$
	for Dimer Model \textbf{1}.
	   }
  \label{FIG3}
\end{figure}

Figures~\ref{FIG3} (a) and (b) show the diagonal VCCs of S$_0$@S$_2$ and S$_1$@S$_2$, respectively.
Furthermore, Figs.~\ref{FIG3} (c) and (d) show the off-diagonal VCCs of 
S$_0$@S$_2$ $\leftarrow$ S$_2$@S$_2$ and S$_1$@S$_2$ $\leftarrow$ S$_2$@S$_2$, respectively.
The diagonal VCCs of S$_1$@S$_2$ are extremely small
where the largest VCC of vibrational mode 233 is $0.79 \times 10^{-4}$ a.u.
In addition, both the off-diagonal VCCs 
of S$_1$@S$_2$ $\leftarrow$ S$_2$@S$_2$ and  S$_0$@S$_2$ $\leftarrow$ S$_2$@S$_2$ are small.
These results indicate that the internal conversions
from S$_2$ to S$_0$ as well as to S$_1$ are suppressed 
while the radiative transition between S$_2$ and S$_0$
is enabled with the large oscillator strength (see Table \ref{TABLE1}).
It should be noted, because of the extremely small diagonal VCCs of S$_1$@S$_2$, that 
the internal conversion from S$_2$ to S$_1$ is suppressed, 
thereby enabling the fluorescence from S$_2$.

\begin{figure*}[h]
\centering
    \includegraphics[width=0.8\hsize]{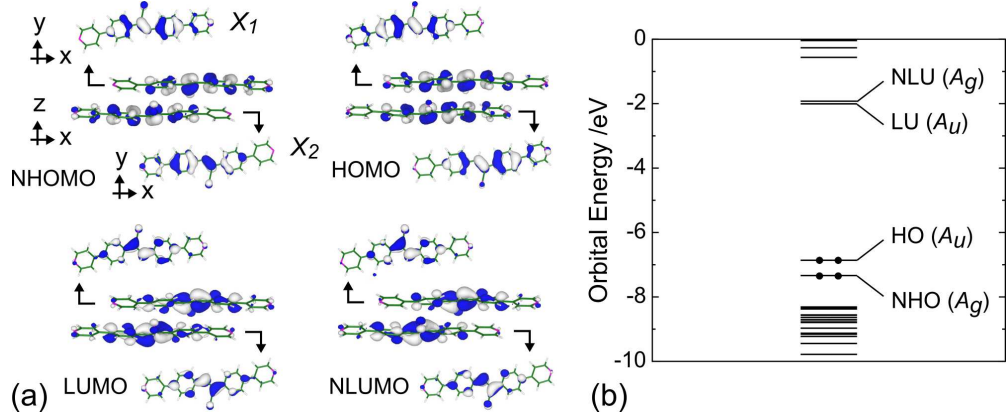}
  \caption{
	(a) Frontier orbitals and 
	(b) orbital levels of Dimer Model \textbf{1} at the S$_2$ optimized structure.
	$X_1$ and $X_2$ are the constituent molecules of Dimer Model \textbf{1}.
	Isosurface values of the frontier orbitals are $3.0\times 10^{-2}$ a.u.
	   }
  \label{FIG4}
\end{figure*}

Figures~\ref{FIG4} (a) and (b) present the frontier orbitals and orbital levels 
at the S$_2$ optimized structure, respectively.
The adiabatic wave functions are delocalized over the molecules,
thereby indicating the excimer formation in solid phase.
Herein, an excimer is defined as an excited state delocalized over a dimer.
The AD S$_2$ is an excimer state without an intermolecular charge transfer character.
The relaxed S$_1$ also forms an excimer state similar to S$_2$.
In the present case, the delocalized electronic states are obtained 
because Dimer Model \textbf{1} belongs to $C_i$ symmetry even in the AD state.
The NHOMO and HOMO as well as the LUMO and NLUMO of the excimer are pseudo-degenerate,
which are approximately expressed as
\begin{eqnarray}	
	\psi_{{\rm NHO}} &\approx& \frac{1}{\sqrt{2}} ( \phi_{{\rm HO}} (X_1) + \phi_{{\rm HO}} (X_2)), \\
	\psi_{{\rm HO}}  &\approx& \frac{1}{\sqrt{2}} ( \phi_{{\rm HO}} (X_1) - \phi_{{\rm HO}} (X_2)), \\
	\psi_{{\rm LU}}  &\approx& \frac{1}{\sqrt{2}} ( \phi_{{\rm LU}} (X_1) - \phi_{{\rm LU}} (X_2)), \\
	\psi_{{\rm NLU}} &\approx& \frac{1}{\sqrt{2}} ( \phi_{{\rm LU}} (X_1) + \phi_{{\rm LU}} (X_2)),
\end{eqnarray}
where $\phi_{{\rm HO/LU}} (X_1/X_2)$
denotes the HOMO/LUMO of molecule $X_1$/$X_2$ comprising the model.
$\phi_{{\rm HO/LU}} (X_2)$ are obtained by a symmetry operation of $\phi_{{\rm HO/LU}} (X_1)$.
The frontier orbitals of the excimer are represented as 
the linear combinations of the HOMOs and LUMOs of the constituents
with opposite signs.

\begin{figure*}[h]
\centering
    \includegraphics[width=0.8\hsize]{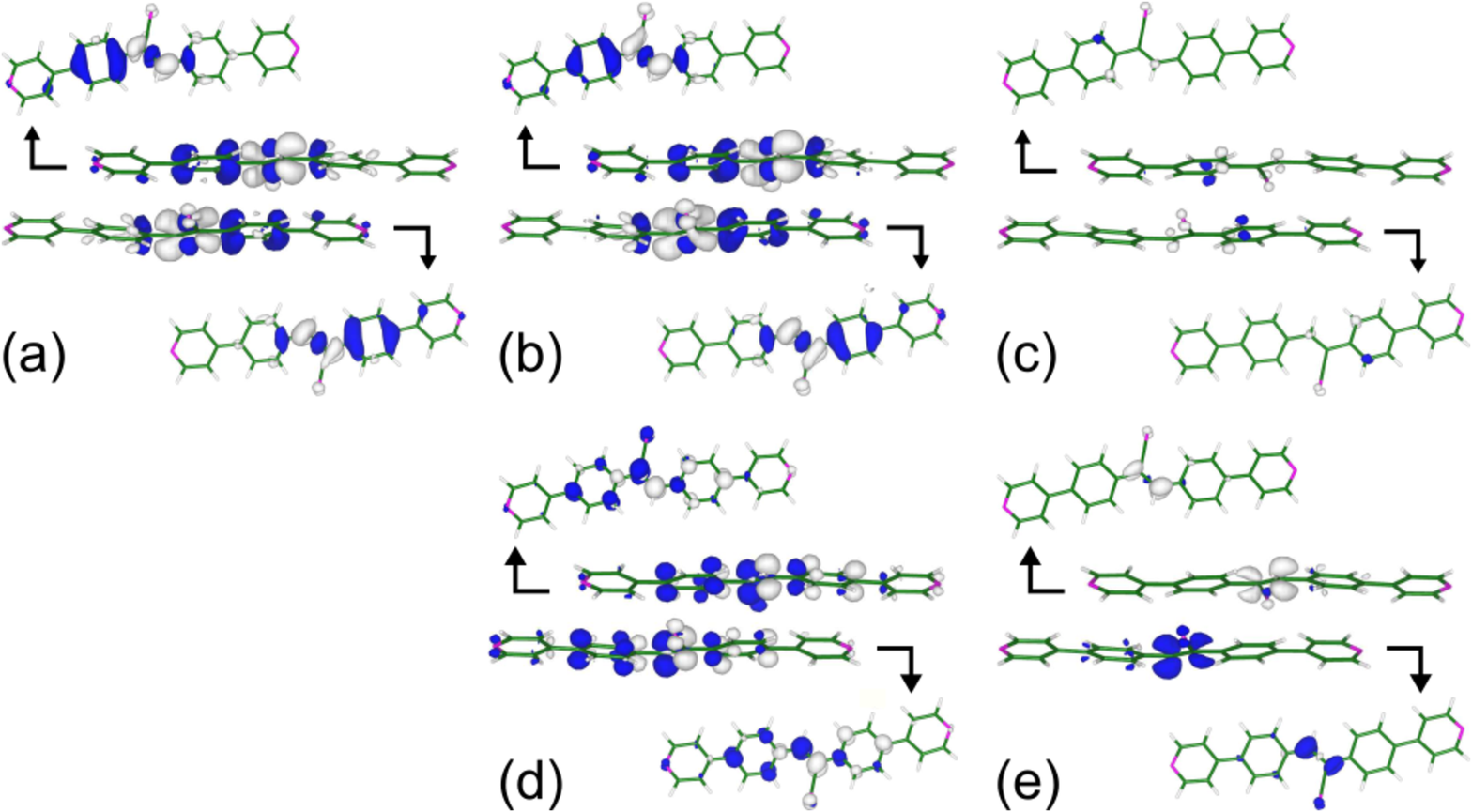}
  \caption{
	Electron density differences of 
	(a) S$_1$@S$_2$-S$_0$@S$_2$, (b) S$_2$@S$_2$-S$_0$@S$_2$, and (c) S$_2$@S$_2$-S$_1$@S$_2$
        of Dimer model \textbf{1}.
	Isosurface values are $1.0\times 10^{-3}$ a.u.
	Overlap densities of (d) S$_2$@S$_2$-S$_0$@S$_2$, and (e) S$_2$@S$_2$-S$_1$@S$_2$
        of Dimer model \textbf{1}.
	Isosurface values are $2.0\times 10^{-3}$ a.u.
	   }
  \label{FIG5}
\end{figure*}

S$_1$ mainly consists of the HOMO-LUMO and NHOMO-NLUMO excited configurations,
whereas S$_2$ mainly consists of the HOMO-NLUMO and NHOMO-LUMO excited configurations
(Table~\ref{TABLE1}).
Since the NHOMO/HOMO and LUMO/NLUMO are pseudo-degenerate,
S$_1$ and S$_2$ are pseudo-degenerate.
Figure~\ref{FIG5} (a) and (b) show the electron density differences of 
S$_2$@S$_2$-S$_0$@S$_2$, $\Delta \rho_{20}$, and S$_1$@S$_2$-S$_0$@S$_2$, $\Delta \rho_{10}$, respectively.
$\Delta \rho_{20}$ and $\Delta \rho_{10}$ exhibit similar distributions.
On the contrary, the electron density difference of S$_2$@S$_2$-S$_1$@S$_2$, $\Delta \rho_{21}$ 
(Fig.~\ref{FIG5} (c))
exhibits an extremely small distribution.
This leads to the small diagonal VCD of S$_1$@S$_2$,
resulting in the small diagonal VCCs of S$_1$@S$_2$.
Figures~\ref{FIG5} (d) and (e) show the overlap densities of S$_2$@S$_2$-S$_0$@S$_2$, $\rho_{20}$,
and S$_2$@S$_2$-S$_1$@S$_2$, $\rho_{21}$, respectively.
$\rho_{21}$ exhibits a smaller distribution than that of $\rho_{20}$.
The small $\rho_{21}$ contributes to 
the small off-diagonal VCCs of S$_1$@S$_2$ $\leftarrow$ S$_2$@S$_2$.
From the distributions of the overlap densities,
the off-diagonal VCCs of S$_0$@S$_2$ $\leftarrow$ S$_2$@S$_2$ could be
much larger than those of S$_1$@S$_2$ $\leftarrow$ S$_2$@S$_2$.
However, the values of these off-diagonal VCCs are comparable (Figs.~\ref{FIG3} (c) and (d))
because $\rho_{20}$ is symmetrically localized on the atoms
while $\rho_{21}$ is localized on the bonds.
In general, an overlap density that is symmetrically localized on atoms 
weakly couples to a potential derivative,
resulting in small off-diagonal VCCs
\cite{Uejima2014_14244}.
Therefore, the off-diagonal VCCs of S$_0$@S$_2$ $\leftarrow$ S$_2$@S$_2$ are not large.

Consequently, the small electron density difference and overlap density
between the pseudo-degenerate S$_1$ and S$_2$ excited states leads to 
the extremely small diagonal VCCs of S$_1$@S$_2$ and the small off-diagonal VCCs of S$_1$@S$_2$$\leftarrow$S$_2$@S$_2$.
We discuss the mechanism of the vanished electron density difference and overlap density 
in the pseudo-degenerate electronic system using the Hubbard model in Sec.~\ref{SEC4-2}.

Below, 
we compare the VCs of Dimer Model \textbf{1} in solid phase
with that of Monomer Model in solution phase.
The reducible representation of vibrational modes 
for the monomer belonging to $C_1$ symmetry contains only $A$ irreducible representations,
and the number of the vibrational modes is
\begin{equation}
	\Gamma_{{\rm vib}} (C_1) = 129 A,
\end{equation}
where all vibrational modes are vibronic active modes
that provide the non-zero diagonal and off-diagonal VCCs.
On the other hand, 
the reducible representation for the dimer belonging to $C_i$ symmetry is decomposed as
\begin{equation}
	\Gamma_{{\rm vib}} (C_i) = 132 A_g + 132 A_u,
\end{equation}
where $A_g$ and $A_u$ are the totally and non-totally symmetric modes, respectively.
The vibronic active mode for the diagonal VC is $A_g$ and that for the off-diagonal VC is $A_u$.
Therefore, the number of the vibronic active modes is almost the same in the monomer and dimer
although the total number of vibrational modes is higher in the dimer.
The numbers of irreducible representations of vibrational modes in excimers 
with $C_i$, $C_2$, and $C_s$ symmetry 
are summarized in Section S2.

\begin{figure}[h]
\centering
    \includegraphics[width=0.8\hsize]{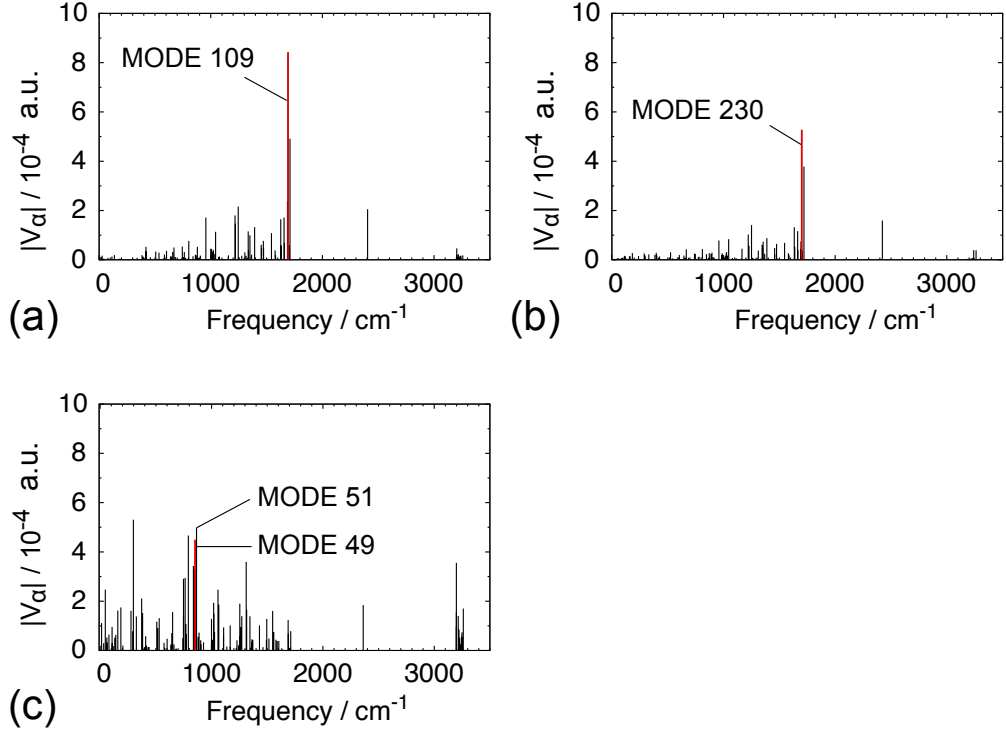}
  \caption{
	Diagonal VCCs (a) of Monomer Model in the FC S$_1$ state
	and (b) of Dimer Model \textbf{1} in the FC S$_2$ state.
	(c) Off-diagonal VCCs of S$_0$@S$_1$ $\leftarrow$ S$_1$@S$_1$ of Monomer Model.
	   }
  \label{FIG6}
\end{figure}

Figures~\ref{FIG6} (a) and (b) show the diagonal VCCs 
of Monomer Model in the FC S$_1$ state and Dimer Model \textbf{1} in the FC S$_2$ state,
respectively.
The diagonal VCCs are greatly reduced due to the excimer formation
where the largest VCC of mode $109$ in Monomer Model is $8.42\times10^{-4}$ a.u.
and that of mode $230$ in Dimer Model \textbf{1} is $5.27\times10^{-4}$ a.u.
This result indicates that 
the internal conversion from S$_2$ to S$_0$ in Dimer Model \textbf{1} is suppressed 
in comparison with the one from S$_1$ to S$_0$ in Monomer Model
because the rate constant of the internal conversion is strongly correlated with the diagonal VCCs
\cite{Uejima2014_14244}.
The off-diagonal VCCs of S$_0$@S$_1$ $\leftarrow$ S$_1$@S$_1$ for Monomer Model are
presented in Fig.~\ref{FIG6} (c).
The off-diagonal VCCs of Dimer Model \textbf{1} (Fig.~\ref{FIG3} (c))
are smaller than those of Monomer Model.
For example, 
the off-diagonal VCC of mode $49$ in Monomer Model is $4.49\times10^{-4}$
and that of mode $111$ in Dimer Model \textbf{1}, corresponding to mode $49$ in Monomer Model,
is $0.89 \times 10^{-4}$ a.u.
The reduction of the off-diagonal VCCs in Dimer Model \textbf{1} also contributes to 
the suppression of the internal conversion.

The off-diagonal VCCs of the maximum coupling modes are comparable
between Monomer Model ($5.30\times10^{-4}$ a.u.) and 
Dimer Model \textbf{1} ($5.33\times10^{-4}$ a.u.).
In contrast,
the diagonal VCC of the maximum coupling mode at $1692$ cm$^{-1}$ for Monomer Model 
($8.42\times10^{-4}$ a.u.)
is larger than that at $1701$ cm$^{-1}$ for Dimer Model \textbf{1} 
($5.27\times10^{-4}$ a.u.).
The differences of electronic energies between S$_1$ and S$_0$ is 3.14 eV for Monomer Model,
and between S$_2$ and S$_0$ is 3.51 eV for Dimer Model \textbf{1}.
Suppose that $V_{mn,\alpha}=5.00\times10^{-4}$ a.u.,
$\omega_{\alpha}=1700$ cm$^{-1}$,
$E_m-E_n=3.0$ eV,
$\sigma=500$ cm$^{-1}$,
$E_{{\rm min}}=-11500$ cm$^{-1}$, and
$E_{{\rm max}}=11500$ cm$^{-1}$ in Eq.~(\ref{Eq:IC2}),
the rate constants of the internal conversion 
considering only a diagonal VCC maximum coupling mode are estimated to be
$1.569\times10^{10}$ s$^{-1}$ for Monomer Model and
$5.651\times10^7$ s$^{-1}$ for Dimer Model \textbf{1}.
The ratio of the rate constants of Monomer Model to Dimer Model \textbf{1} is in the order of $10^2$.
These are in good agreement with experiments 
($>1.0\times10^{10}$ s$^{-1}$ in CH$_2$Cl$_2$ solution and
$5.0\times10^7$ s$^{-1}$ in solid phase)
\cite{Nishio2014_686}.
Therefore, the reduction of the diagonal VCCs 
mainly contributes to the suppression of the internal conversion in the solid phase.
It should be noted however that
the rate constant depends 
on the broadening of the density of states $\sigma$
(Table S4)
\cite{Zhang2019_264}.
The broadening of the density of states occurs 
by the interactions with the surrounding environment and vibronic couplings.
Since the evaluation of these effects is out of scope in the present study,
$\sigma$ was treated as an external parameter.
We previously adopted the values of $\sigma$ ranging from 300 to 1200 cm$^{-1}$
to compute the absorption spectra that reproduces the experimental results
\cite{Uejima2014_80,Shigemitsu2012_9100}.
We employed $\sigma=500$ cm$^{-1}$ in estimating the rate constants,
and obtained the good agreement of the experimental values.
Even when different $\sigma$ is used,
the ratios of the rate constants of Monomer Model to Dimer Model \textbf{1} 
do not drastically change (Table S4).

\begin{figure*}[h]
\centering
    \includegraphics[width=0.8\hsize]{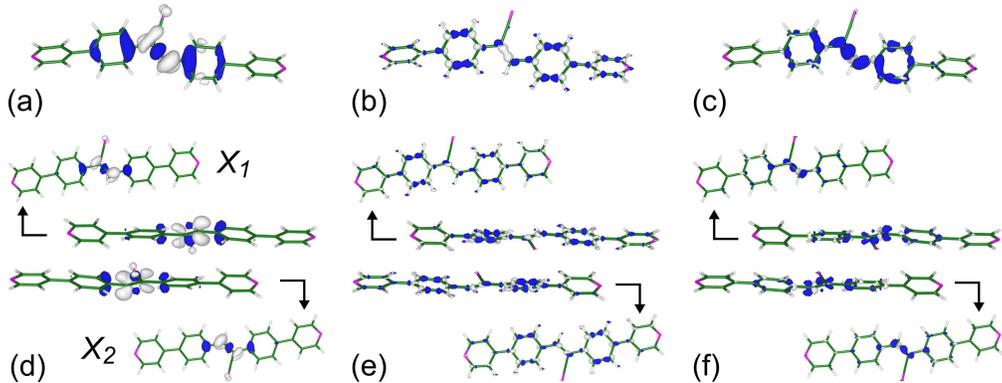}
  \caption{
	(a) Electron density difference, $\Delta \rho_{nm} (\textbf{\textit{x}})$, 	
	(b) potential derivative, $v_{\alpha} (\textbf{\textit{x}})$,
	and (c) diagonal VCD, $\eta_{n, \alpha} (\textbf{\textit{x}})$, of Monomer Model in the FC S$_1$ state
	for the maximum coupling mode (mode 109).
        (d) $\Delta \rho_{nm} (\textbf{\textit{x}})$, (e) $v_{\alpha} (\textbf{\textit{x}})$, and
	(f) $\eta_{n, \alpha} (\textbf{\textit{x}})$ of Dimer Model \textbf{1} in the FC S$_2$ state 
	for the maximum coupling mode (mode 230).
	Isosurface values of $\Delta \rho_{nm}(\textbf{\textit{x}})$, $v_{\alpha} (\textbf{\textit{x}})$, 
	and $\eta_{n,\alpha}(\textbf{\textit{x}})$ are 
	$2.0\times10^{-3}$, $1.0\times10^{-2}$, and $1.0\times10^{-5}$ a.u., respectively.
	   }
  \label{FIG7}
\end{figure*}

To determine the origin of the VCCs, 
VCD analyses are performed.
Figure~\ref{FIG7} shows the results of the diagonal VCD analyses for the maximum coupling mode
of Monomer Model and Dimer Model \textbf{1}.
$\Delta \rho_{nm} (\textbf{\textit{x}})$ of Monomer Model 
strongly couples with $v_{\alpha}(\textbf{\textit{x}})$ of mode 109,
leading to a large $\eta_{n,\alpha} (\textbf{\textit{x}})$ 
that is particularly localized on the C=C bond.
Thus, the spatial integration of $\eta_{n,\alpha} (\textbf{\textit{x}})$ of mode 109 
provides the largest VCC.
$\Delta \rho_{nm} (\textbf{\textit{x}})$, $v_{\alpha} (\textbf{\textit{x}})$, 
and $\eta_{n,\alpha} (\textbf{\textit{x}})$ 
of Dimer Model \textbf{1} are delocalized over the molecules.
In general, delocalized $\Delta \rho_{nm} (\textbf{\textit{x}})$ and $v_{\alpha} (\textbf{\textit{x}})$ 
yield smaller diagonal VCCs than localized ones
\cite{Shizu2013_215}.
$\Delta \rho_{nm} (\textbf{\textit{x}})$ of $X_1$ (or $X_2$)
exhibits a similar distribution to that of Monomer Model.
The value of $\Delta \rho_{nm} (\textbf{\textit{x}})$ of $X_1$ 
is one half of that of Monomer Model
because the spatial integration of $\Delta \rho_{nm} (\textbf{\textit{x}})$ is zero by definition.
In addition, 
$v_{\alpha}(\textbf{\textit{x}})$ of $X_1$ is
$1/\sqrt{2}$ times that of Monomer Model 
owing to the normalized condition of vibrational modes.
Since $\eta_{n,\alpha}(\textbf{\textit{x}})$ is expressed 
as the product of $\Delta \rho_{nm} (\textbf{\textit{x}})$ and $v_{\alpha}(\textbf{\textit{x}})$,
the diagonal VCD of $X_1$ is $1/(2\sqrt{2})$ times that of Monomer Model.
Therefore, the diagonal VCCs of Dimer Model \textbf{1}, obtained by the spatial integration of the diagonal VCD,
are $1/\sqrt{2}$ times those of the Monomer Model.
The ratio of the largest diagonal VCC of Dimer Model \textbf{1} to Monomer Model is $0.626$,
which is approximately equal to $1/\sqrt{2} \approx 0.707$;
however deviation occurs 
because the structures of $X_1$ and the monomer are not the same.

In Dimer Model \textbf{1}, 
the reduction of the diagonal VCCs results from the delocalized electronic states.
In contrast, in Dimer Model \textbf{2}
the electronic states are localized on a single molecule in the adiabatic excited states
(Fig. S3),
which causes the properties of excited states to be similar to those of the monomer.
Therefore, the reduction of the diagonal VCCs is not expected in Dimer Model \textbf{2}.
The electronic states are delocalized over molecules 
in Dimer Model \textbf{3}
(Fig. S4),
suggesting small diagonal VCCs as in the case for Dimer Model \textbf{1}.

\begin{figure}[h]
\centering
    \includegraphics[width=0.4\hsize]{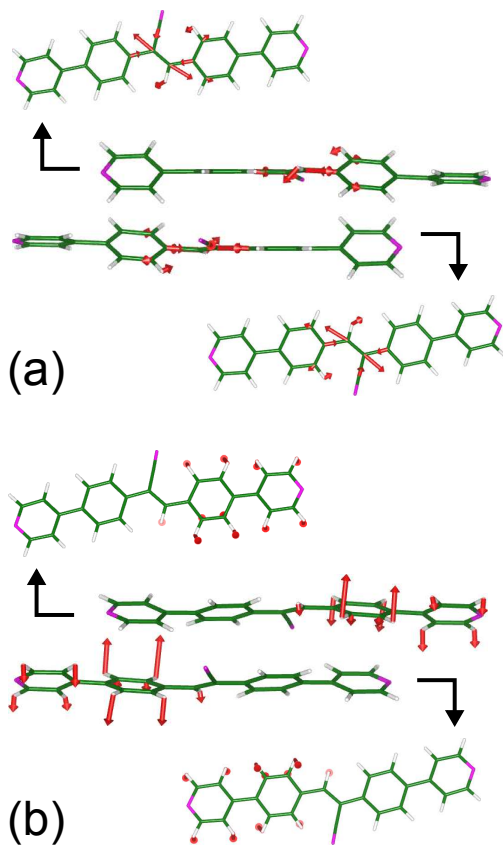}
  \caption{
	(a) Effective mode for the excimer formation of Dimer Model \textbf{1} in the FC S$_2$ state.
	This is an intramolecular vibration.
	(b) Vibrational mode 111 of Dimer Model \textbf{1} in the AD S$_2$ state;
	the off-diagonal VCC of this mode is decreased due to the packing effect.
	This is an intermolecular vibration.
	   }
  \label{FIG8}
\end{figure}

The driving force of the excimer formation is the diagonal VCCs.
Figure~\ref{FIG8} (a) shows the effective mode for the excimer formation 
of Dimer Model \textbf{1} in the FC S$_2$ state.
The effective mode is defined by the sum of normal modes weighted by the diagonal VCCs:
\begin{equation}
	\xi = \sum_{\alpha} \frac{V_{n,\alpha}}{\sqrt{ \sum_{\alpha} |V_{n,\alpha}|^2}} Q_{\alpha},
\end{equation}
which is the steepest descent direction of the vibrational relaxation
\cite{Sato2012_257}.
The effective mode is the intramolecular vibration rather than the intermolecular one.
This indicates that the excimer formation is induced by the intramolecular vibration.

\begin{figure}[h]
\centering
    \includegraphics[width=0.8\hsize]{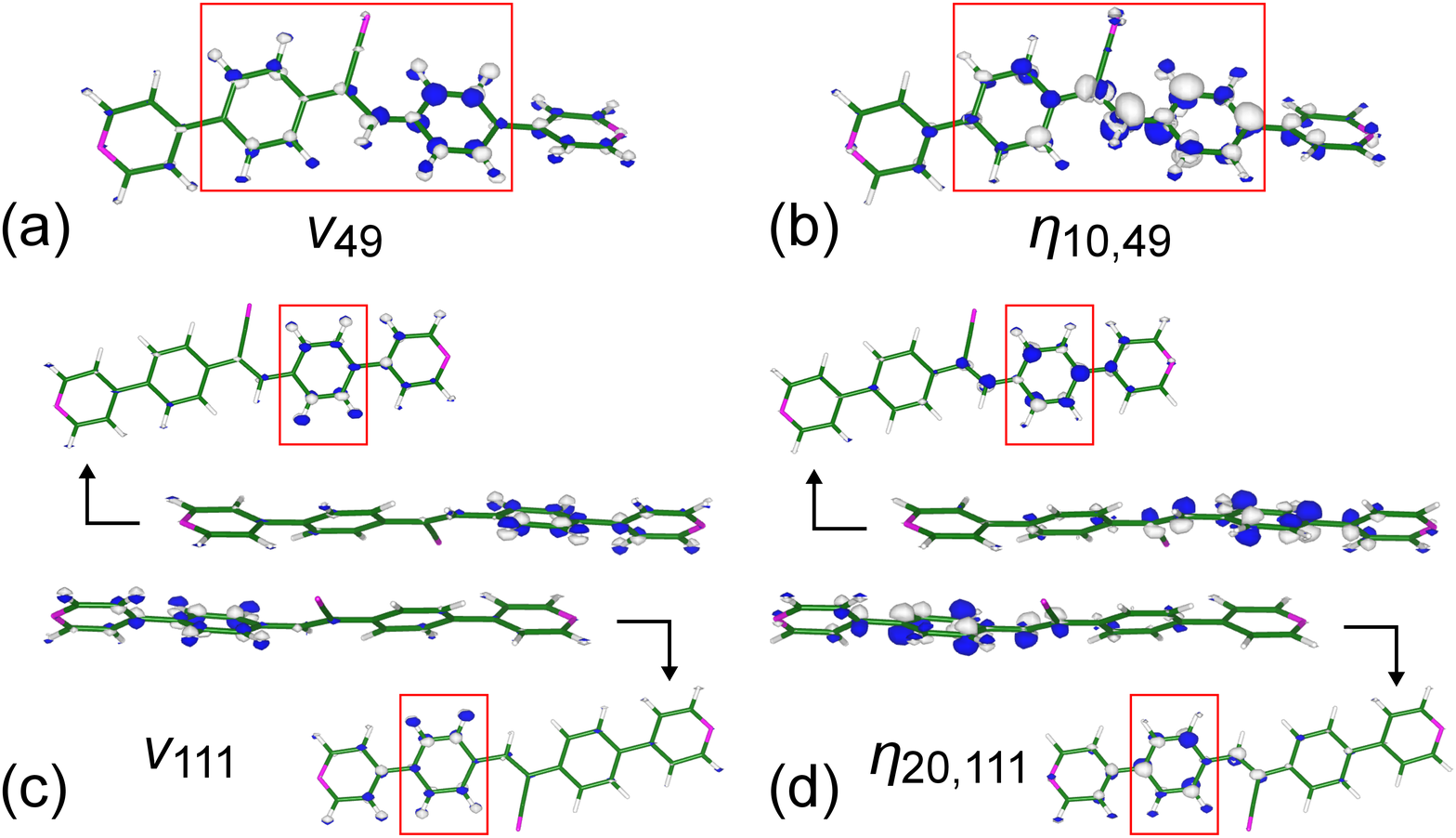}
  \caption{
	(a) Potential derivative, $v_{\alpha} (\textbf{\textit{x}})$,
	and (b) off-diagonal VCD, $\eta_{mn, \alpha} (\textbf{\textit{x}})$, of 
	S$_1$@S$_1$-S$_0$@S$_1$ of Monomer Model.
	(c) $v_{\alpha} (\textbf{\textit{x}})$
	and (b) $\eta_{mn, \alpha} (\textbf{\textit{x}})$ of 
	S$_2$@S$_2$-S$_0$@S$_2$ of Dimer Model \textbf{1}.
	Isosurface values of $v_{\alpha} (\textbf{\textit{x}})$ and $\eta_{mn,\alpha}(\textbf{\textit{x}})$ are 
	$1.0\times10^{-2}$ and $1.0\times10^{-5}$ a.u., respectively.
	   }
  \label{FIG9}
\end{figure}

Figure~\ref{FIG9} shows the results of the off-diagonal VCD analyses
of mode $49$ for Monomer Model and mode $111$ for Dimer Model \textbf{1}.
$v_{49} (\textbf{\textit{x}})$ of Monomer Model is distributed over the stilbene unit.
On the contrary, $v_{111} (\textbf{\textit{x}})$ of Dimer Model \textbf{1} is localized 
on one side of the stilbene unit.
This is because vibrational mode $111$ is the intermolecular vibration (Fig.~\ref{FIG8} (b)),
i.e. torsional mode of stilbene unit,
that is restricted by the surrounding molecules in the solid phase.
In other words, the localization of $v_{\alpha} (\textbf{\textit{x}})$ arises from the packing effect.
As a result, $\eta_{mn,\alpha} (\textbf{\textit{x}})$ of Dimer Model \textbf{1} is 
also localized on one side of the stilbene unit giving the small off-diagonal VCCs.
The degree of localization of $v_{\alpha}(\textbf{\textit{x}})$ 
is expected to depend on the types of vibrational modes.
The off-diagonal VCC of mode $51$ in Monomer Model is $4.98\times10^{-4}$ a.u. (Fig.~\ref{FIG6} (c)),
which is comparable to that of mode $113$ in Dimer Model \textbf{1}, $5.34\times10^{-4}$ a.u.,
(Fig.~\ref{FIG3} (c)).
Figure S5 shows $v_{51}(\textbf{\textit{x}})$ of Monomer Model
and $v_{113} (\textbf{\textit{x}})$ of Dimer Model \textbf{1}.
$v_{113}(\textbf{\textit{x}})$ of Dimer Model \textbf{1}
is distributed over the stilbene unit rather than localized on one side of the stilbene unit
in a similar manner to $v_{51} (\textbf{\textit{x}})$ of Monomer Model.
This is because vibrational mode 113 is the intramolecular vibration 
that is not easily affected by the surrounding molecules
(Fig. S5).
Therefore, the off-diagonal VCCs corresponding to these modes are not reduced by the packing effect.
The packing effect in solid phase is visualized by the potential derivative.

The small diagonal VCCs contribute to the suppression not only of the internal conversion 
but also of the vibrational relaxation
\cite{Uejima2014_14244}.
The reorganization energies due to vibrational relaxation were calculated to be
$\Delta E$ = 0.582 eV for Monomer Model and 
$\Delta E$ = 0.275 eV for Dimer Model \textbf{1}. 
For Monomer Model,
since the potential energy surfaces of low-frequency torsional modes
were not approximated to harmonic potentials,
the reorganization energies for these vibrational modes 
were evaluated from the potential energy surfaces 
\cite{Uejima2014_14244}
instead of using Eq.~(\ref{Eq:DeltaE})
(Fig. S6).
The reorganization energy of Dimer Model \textbf{1} is smaller than that of Monomer Model,
and suppressed vibrational relaxation is expected in Dimer Model \textbf{1}.

Furthermore, we calculated the electronic states of Decamer Model for the CNPPE solid
(Fig. S7).
Decamer Model has a large oscillator strength in the FC S$_8$ state (Table S7),
and the S$_8$ exciton is expected to be generated by absorption.
The electron density difference between the FC S$_8$ and S$_0$ states 
is delocalized as in the case for the combination of Dimer Model \textbf{1} and \textbf{2} 
(Fig. S8).
The electronic states of Dimer Model \textbf{2} are localized on a single molecule
after vibrational relaxation.
Therefore, Decamer Model can be reduced to Dimer Model \textbf{1} in the adiabatic excited states.
Thus, the modeling of the CNPPE solid using Dimer Model \textbf{1} is considered to be reasonable.

We also performed the calculations of Dimer Model \textbf{1} 
using the long-range corrected $\omega$B97X-D functional
\cite{Chai2008_6615}
in order to check the robustness of the results.
S$_2$ ($A_u$) consisting of the HOMO-1-LUMO and HOMO-LUMO+1 excited configurations
is symmetry-allowed from the Laport\'{e} rule (Table S8).
In addition, the frontier orbitals in the AD S$_2$ state are delocalized over the molecules (Fig. S9),
and the excited states do not have a charge transfer character.
Therefore, the same conclusions as using the M06-2X functional can be deduced.

\subsection{Hubbard model in a pseudo-degenerate electronic system\label{SEC4-2}}
\begin{figure}[h]
\centering
    \includegraphics[width=0.4\hsize]{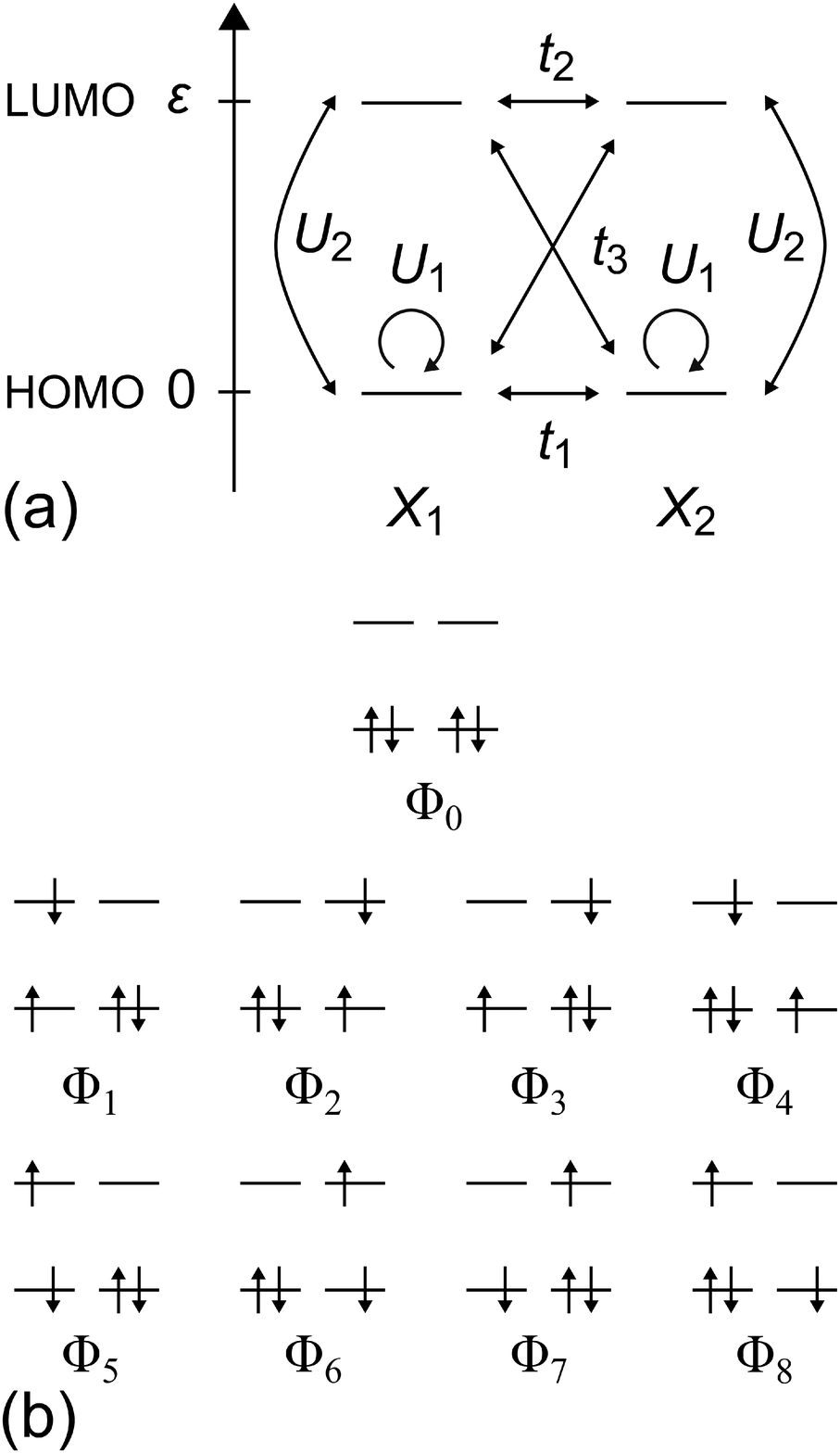}
  \caption{
	(a) Hubbard model of a system consisting of $X_1$ and $X_2$.
	$\epsilon$ is the energy gap between the HOMO and LUMO.
	$t_1$, $t_2$, and $t_3$ are the hopping integrals for the HOMO-HOMO, LUMO-LUMO, and HOMO-LUMO,
	respectively, between $X_1$ and $X_2$.
	$U_1$ and $U_2$ are the Coulomb interactions for the HOMO-HOMO and HOMO-LUMO, respectively,
	within $X_1$ or $X_2$.
	(b) Electronic ground and excited configurations assuming a single excitation.
	   }
  \label{FIG10}
\end{figure}

In this section, we discuss the disappearance mechanism of 
the electron density difference and overlap density based on the Hubbard model.
We consider an excited electronic structure of a system 
consisting of the same molecules $X_1$ and $X_2$
(Fig.~\ref{FIG10} (a)).
Each molecule is not necessarily symmetric,
but the dimer is assumed to have a symmetry such as $C_i$, $C_2$, or $C_s$.
For simplicity, only the HOMOs and LUMOs of $X_1$ and $X_2$ are considered.
The energy gap between the HOMO and LUMO is expressed as $\epsilon$.
The hopping integrals for the HOMO-HOMO, LUMO-LUMO, and HOMO-LUMO between $X_1$ and $X_2$
are denoted by $t_1$, $t_2$, and $t_3$, respectively.
The Coulomb interactions
for the HOMO-HOMO and HOMO-LUMO within $X_1$ or $X_2$ are denoted by 
$U_1$ and $U_2$, respectively.
Considering only a single excitation
as done by the TD-DFT calculations,
there are 9 electronic configurations (Fig.~\ref{FIG10} (b))
where $\ket{\Phi_0}$ is the ground configuration,
$\ket{\Phi_1}$, $\ket{\Phi_2}$, $\ket{\Phi_5}$, and $\ket{\Phi_6}$ are locally-excited configurations,
and $\ket{\Phi_3}$, $\ket{\Phi_4}$, $\ket{\Phi_7}$, and $\ket{\Phi_8}$ 
are charge-transfer configurations.
$t_1$, $t_2$, and $t_3$ are considered to be small compared with $U_1$ and $U_2$
because $t_1$, $t_2$, and $t_3$ are the interactions between $X_1$ and $X_2$
separated from each other 
while $U_1$ and $U_2$ are the interactions within $X_1$ or $X_2$.
The model Hamiltonian for the basis defined in Fig.~\ref{FIG10} (b) is given by
\begin{equation}
\scalebox{0.55}{$
\begin{pmatrix}
2U_1 & 0                & 0                & t_3               & t_3               & 0                & 0                & t_3              & t_3 \\
     & \epsilon+U_1+U_2 & 0                & t_2               & t_1               & 0                & 0                & 0                & 0   \\
     &                  & \epsilon+U_1+U_2 & t_1               & t_2               & 0                & 0                & 0                & 0   \\
     &                  &                  & \epsilon+U_1+2U_2 & 0                 & 0                & 0                & 0                & 0   \\
     &                  &                  &                   & \epsilon+U_1+2U_2 & 0                & 0                & 0                & 0    \\
     &                  &                  &                   &                   & \epsilon+U_1+U_2 & 0                & t_2              & t_1 \\
     &                  &                  &                   &                   &                  & \epsilon+U_1+U_2 & t_1              & t_2 \\
     &                  &                  &                   &                   &                  &                  &\epsilon+U_1+2U_2 & 0   \\
     &                  &                  &                   &                   &                  &                  &                  & \epsilon+U_1+2U_2   \\
\end{pmatrix}.
$}
\end{equation}
Employing the Rayleigh--Schr\"odinger perturbation theory 
considering $t_1$, $t_2$, and $t_3$ as perturbations,
the electronic states of the excited states are expressed as
\begin{eqnarray}
        \ket{\Psi^s_1}  
        & = & \frac{1}{2} ( \ket{\Phi_1} + \ket{\Phi_5} + \ket{\Phi_2} + \ket{\Phi_6} ) \notag \\
        & - & \frac{t_1+t_2}{2 U_2} ( \ket{\Phi_3} + \ket{\Phi_7} + \ket{\Phi_4} + \ket{\Phi_8} ), \\
        \ket{\Psi^s_2}  
        & = & \frac{1}{2} (\ket{\Phi_3} + \ket{\Phi_7} + \ket{\Phi_4} + \ket{\Phi_8}) \notag \\
        & + & \frac{t_1+t_2}{2 U_2} ( \ket{\Phi_1} + \ket{\Phi_5} + \ket{\Phi_2} + \ket{\Phi_6} ), \\
        \ket{\Psi^a_1}  
        & = & \frac{1}{2} ( \ket{\Phi_1} + \ket{\Phi_5} - \ket{\Phi_2} - \ket{\Phi_6} ) \notag \\
        & - & \frac{t_2-t_1}{2 U_2} ( \ket{\Phi_3} + \ket{\Phi_7} - \ket{\Phi_4} - \ket{\Phi_8} ), \\
        \ket{\Psi^a_2} 
        & = & \frac{1}{2} (\ket{\Phi_3} + \ket{\Phi_7} - \ket{\Phi_4} - \ket{\Phi_8}) \notag \\
        & + & \frac{t_2-t_1}{2 U_2} ( \ket{\Phi_1} + \ket{\Phi_5} - \ket{\Phi_2} - \ket{\Phi_6} ),
\end{eqnarray}
where $\ket{\Psi^s_1}$ and $\ket{\Psi^s_2}$ are symmetric and 
$\ket{\Psi^a_1}$ and $\ket{\Psi^a_2}$ are antisymmetric electronic states.
$\ket{\Psi^s_1}$ and $\ket{\Psi^s_2}$ belong to 
different symmetries from $\ket{\Psi^a_1}$ and $\ket{\Psi^a_2}$.
The electronic states of the ground state is expressed as
\begin{equation}
	\ket{\Psi_0} 
	= \ket{\Phi_0} 
	- \frac{t_3}{\epsilon-U_1+U_2} ( \ket{\Phi_3} + \ket{\Phi_7} + \ket{\Phi_4} + \ket{\Phi_8}),
\end{equation}
which depends on $t_3$ unlike the electronic states of the excited states.

Employing the Rayleigh--Schr\"odinger perturbation theory,
the energies of the ground and excited states are given by
\begin{eqnarray}
	E_0 & = & 2U_1-\frac{4 t_3^2}{\epsilon-U_1+2U_2}, \\
	E_1^s & = & \epsilon+U_1+U_2-\frac{(t_1+t_2)^2}{U_2},  \\
	E_2^s & = & \epsilon+U_1+2U_2+\frac{(t_1+t_2)^2}{U_2},\\
	E_1^a & = & \epsilon+U_1+U_2-\frac{(t_2-t_1)^2}{U_2}, \\ 
	E_2^a & = & \epsilon+U_1+2U_2+\frac{(t_2-t_1)^2}{U_2}.
\end{eqnarray}
The energy difference between $E_1^s$ and $E_1^a$ as well as $E_2^s$ and $E_2^a$
is $|4 t_1 t_2/U_2|$.
Since $4t_1 t_2/U_2$ is assumed to be small,
$\ket{\Psi^s_1}$ and $\ket{\Psi^a_1}$ as well as $\ket{\Psi^s_2}$ and $\ket{\Psi^a_2}$ are pseudo-degenerate.
The orders of the excited states are
$E_1^s \lesssim E_1^a < E_2^a \lesssim E_2^s$ for $t_1 t_2 > 0$
and $E_1^a \lesssim E_1^s < E_2^s \lesssim E_2^a$ for $t_1 t_2 < 0$: 
i.e. the symmetry of excited states changes 
according to the sign of the product $t_1 t_2$ determined by the alignment of molecules.
The excited state degenerates for $t_1 t_2 = 0$.
When $t_1$, $t_2$, and $t_3$ are no longer small compared to $U_1$ and $U_2$,
namely the perturbation theory is not valid,
the degeneracy occurs in the case of $t_1 \neq 0$ or $t_2 \neq 0$
(see Fig. S10).

When the system has $C_i$ symmetry and $t_1 t_2 > 0$,
S$_1$ is symmetry-forbidden $A_g$ and S$_2$ is symmetry-allowed $A_u$.
This is the case for Dimer Model \textbf{1} of the CNPPE solid.
The electron density differences between S$_1$ ($\ket{\Psi_1^s}$) and S$_0$ ($\ket{\Psi_0}$) 
as well as S$_2$ ($\ket{\Psi_1^a}$) and S$_0$ are given by
\begin{eqnarray}
	\Delta \rho_{20} & = & \Delta \rho_{10}  \notag \\
	& = & 
	\left\{ ( |\phi_{{\rm LU}} (X_1)|^2 - |\phi_{{\rm HO}} (X_1)|^2) -
	        ( |\phi_{{\rm LU}} (X_2)|^2 - |\phi_{{\rm HO}} (X_2)|^2) \right\}. \notag \\
\end{eqnarray}
Thus, both $\Delta \rho_{20}$ and $\Delta \rho_{10}$ are expressed as 
the difference between the LUMOs and HOMOs of $X_1$ and $X_2$.
In contrast, the electron density difference between S$_2$ and S$_1$ is expressed as
\begin{equation}
\Delta \rho_{21} = \Delta \rho_{20} - \Delta \rho_{10} = 0.
\end{equation}
Therefore, $\Delta \rho_{21}$ is cancelled 
because $\Delta \rho_{20}$ and $\Delta \rho_{10}$ are the same 
due to the pseudo-degeneracy of S$_1$ and S$_2$.
The overlap density between S$_2$ and S$_0$ as well as between S$_2$ and S$_1$ are given by
\begin{eqnarray}
	\rho_{20} & = & 
	\frac{1}{\sqrt{2}} \left( \phi_{{\rm HO}} (X_1) \phi_{{\rm LU}}(X_1) 
	                        - \phi_{{\rm HO}} (X_2) \phi_{{\rm LU}}(X_2) \right), \\
	\rho_{21} & = & \frac{1}{2} 
	\left\{ ( |\phi_{{\rm LU}} (X_1)|^2 - |\phi_{{\rm HO}} (X_1)|^2) -
	        ( |\phi_{{\rm LU}} (X_2)|^2 - |\phi_{{\rm HO}} (X_2)|^2) \right\} \notag \\
	& \approx & 0. 
	\label{Eq:RHOS2-S1}
\end{eqnarray}
$\rho_{20}$ is expressed as the non-vanishing product of HOMOs and LUMOs.
In contrast, $\rho_{21}$ is cancelled 
because it is expressed as the difference between the square of the HOMOs and LUMOs.
The other pairs of electron density differences and overlap densities 
in the pseudo-degenerate electronic system are
summarized in Table S9 and Table S10, respectively.

Kasha originally discussed the spectroscopic properties of aggregates (H- and J-aggregations)
using the long-range Coulomb couplings alone
\cite{Kasha1963_55}.
Spano \textit{et al.} extended the Kasha's model 
by adding the short-range charge-transfer mediated excitonic couplings
\cite{Hestand2017_341,Hestand2018_7069}.
This model explains that, when the short-range couplings, or intermolecular hopping integrals, are dominant,
excited states properties are determined from the relative phases of the overlap between adjacent molecules,
and the fluorescence could occur even in the H-aggregation.
In the present system, however, the intermolecular hopping integrals are not dominant 
to apply the short-range coupling model.

\section{Concluding remarks}

The origin of AIEE in the CNPPE solid 
was investigated by the ONIOM method using the TD-DFT calculations.
The pseudo-degeneracy arising from the excimer formation in the solid phase
gives the vanished electron density difference and overlap density 
between S$_1$ and S$_2$,
which suggests that the fluorescence from the second excited state was possible
against Kasha's rule 
because the transitions from the second to first excited states are suppressed.
The electronic states delocalized over the molecules
reduce the diagonal VCCs in the solid phase to 
approximately $1/\sqrt{2}$ times those in solution phase.
In addition, the packing effect in the solid phase reduces the off-diagonal VCCs 
of the intermolecular vibrations.
These results indicate that the internal conversions from excited to ground states 
are more suppressed in the solid phase than in the solution phase.
The molecular orientation affecting the relative signs of the intermolecular hopping integrals
plays an important role in determining the pseudo-degenerate excited states properties.
When the product of the hopping integrals for HOMOs and LUMOs is negative,
the first excited state is fluorescent (J-type aggregation).
In contrast, the product of the hopping integrals is positive,
the first excited state is dark and the second one is fluorescent (H-type aggregation).
Whether the first or second excited states are fluorescent,
the AIEE could occur as long as the excimer forms in the aggregation phases
because the excimer formation gives rise to the decrease of the diagonal VCCs.

In this study, we discussed a dimer with $C_i$ site symmetry.
Other cyano-substituted compounds exhibiting AIEE, such as
cyano-substituted bis(4-bromophenyl)-fumaronitrile,
bis(3-trifluoromethylphenyl)fumaronitrile,
bis(4-methoxyphenyl)-fumaronitrile
\cite{Yeh2004_6455}, and
cyano-substituted oligo(\textit{para}-phenylene vinylene) (CN-DPDSB)
\cite{Li2007_231}
also have $C_i$ site symmetry in their crystal structures,
which suggests that AIEE may occur due to the pseudo-degeneracy in these compounds.
It should be noted, however, that
other symmetries including $C_1$ could be also possible for the appearance of AIEE
if the pseudo-degenerate electronic states are generated in the aggregation phases.

A molecule which forms excimers with delocalized excited electronic states can be
fluorescent in the aggregation phases, 
even if the molecule is not fluorescent in an isolated state, such as in solution or vacuum. 
Accordingly, we can obtain the following design principle for AIEE:
\textit{
a candidate molecule for AIEE should have pseudo-degenerate adiabatic electronic states 
in the aggregation phases originating from the excimer formation.
}

\begin{acknowledgement}
This study was supported by JSPS KAKENHI Grant Number JP17H05259 
in Scientific Research on Innovative Areas "Photosynergetics" 
and by Element Strategy Initiative of MEXT, Grant Number JPMXP0112101003.
Numerical calculations were partly performed at
Supercomputer System, Institute for Chemical Research, Kyoto University,
Academic Center for Computing and Media Studies (ACCMS), Kyoto University,
and Research Center for Computational Science, Okazaki.
\end{acknowledgement}

\begin{suppinfo}
\end{suppinfo}

\bibliography{refs}

\providecommand{\latin}[1]{#1}
\providecommand*\mcitethebibliography{\thebibliography}
\csname @ifundefined\endcsname{endmcitethebibliography}
  {\let\endmcitethebibliography\endthebibliography}{}
\begin{mcitethebibliography}{41}
\providecommand*\natexlab[1]{#1}
\providecommand*\mciteSetBstSublistMode[1]{}
\providecommand*\mciteSetBstMaxWidthForm[2]{}
\providecommand*\mciteBstWouldAddEndPuncttrue
  {\def\EndOfBibitem{\unskip.}}
\providecommand*\mciteBstWouldAddEndPunctfalse
  {\let\EndOfBibitem\relax}
\providecommand*\mciteSetBstMidEndSepPunct[3]{}
\providecommand*\mciteSetBstSublistLabelBeginEnd[3]{}
\providecommand*\EndOfBibitem{}
\mciteSetBstSublistMode{f}
\mciteSetBstMaxWidthForm{subitem}{(\alph{mcitesubitemcount})}
\mciteSetBstSublistLabelBeginEnd
  {\mcitemaxwidthsubitemform\space}
  {\relax}
  {\relax}

\bibitem[Hong \latin{et~al.}(2009)Hong, Lam, and Tang]{Hong2009_4332}
Hong,~Y.; Lam,~J. W.~Y.; Tang,~B.~Z. Aggregation-Induced Emission: Phenomenon,
  Mechanism and Applications. \emph{Chem. Commun.} \textbf{2009},
  4332--4353\relax
\mciteBstWouldAddEndPuncttrue
\mciteSetBstMidEndSepPunct{\mcitedefaultmidpunct}
{\mcitedefaultendpunct}{\mcitedefaultseppunct}\relax
\EndOfBibitem
\bibitem[Hong \latin{et~al.}(2011)Hong, Lam, and Tang]{Hong2011_5361}
Hong,~Y.; Lam,~J. W.~Y.; Tang,~B.~Z. Aggregation-Induced Emission. \emph{Chem.
  Soc. Rev.} \textbf{2011}, \emph{40}, 5361--5388\relax
\mciteBstWouldAddEndPuncttrue
\mciteSetBstMidEndSepPunct{\mcitedefaultmidpunct}
{\mcitedefaultendpunct}{\mcitedefaultseppunct}\relax
\EndOfBibitem
\bibitem[Mei \latin{et~al.}(2014)Mei, Hong, Lam, Qin, Tang, and
  Tang]{Mei2014_5429}
Mei,~J.; Hong,~Y.; Lam,~J. W.~Y.; Qin,~A.; Tang,~Y.; Tang,~B.~Z.
  Aggregation-Induced Emission: the Whole is More Brilliant than the Parts.
  \emph{Adv. Mater.} \textbf{2014}, \emph{26}, 5429--5479\relax
\mciteBstWouldAddEndPuncttrue
\mciteSetBstMidEndSepPunct{\mcitedefaultmidpunct}
{\mcitedefaultendpunct}{\mcitedefaultseppunct}\relax
\EndOfBibitem
\bibitem[Chen \latin{et~al.}(2003)Chen, Law, Lam, Dong, Lo, Williams, Zhu, and
  Tang]{Chen2003_1535}
Chen,~J.; Law,~C. C.~W.; Lam,~J. W.~Y.; Dong,~Y.; Lo,~S. M.~F.;
  Williams,~I.~D.; Zhu,~D.; Tang,~B.~Z. Synthesis, Light emission,
  Nanoaggregation, and Restricted Intramolecular Rotation of 1,1-Substituted
  2,3,4,5-Tetraphenylsiloles. \emph{Chem. Mater.} \textbf{2003}, \emph{15},
  1535--1546\relax
\mciteBstWouldAddEndPuncttrue
\mciteSetBstMidEndSepPunct{\mcitedefaultmidpunct}
{\mcitedefaultendpunct}{\mcitedefaultseppunct}\relax
\EndOfBibitem
\bibitem[Yu \latin{et~al.}(2005)Yu, Yin, Liu, Chen, Xu, Sun, Ma, Zhan, Peng,
  Shuai, Tang, Zhu, Fang, and Luo]{Yu2005_6335}
Yu,~G.; Yin,~S.; Liu,~Y.; Chen,~J.; Xu,~X.; Sun,~X.; Ma,~D.; Zhan,~X.;
  Peng,~Q.; Shuai,~Z.; Tang,~B.; Zhu,~D.; Fang,~W.; Luo,~Y. Structures,
  Electronic states, Photoluminescence, and Carrier Transport Properties of
  1,1-Disubstituted 2,3,4,5-Tetraphenylsiloles. \emph{J. Am. Chem. Soc.}
  \textbf{2005}, \emph{127}, 6335--6346\relax
\mciteBstWouldAddEndPuncttrue
\mciteSetBstMidEndSepPunct{\mcitedefaultmidpunct}
{\mcitedefaultendpunct}{\mcitedefaultseppunct}\relax
\EndOfBibitem
\bibitem[Zhao \latin{et~al.}(2015)Zhao, He, and Tang]{Zhao2015_5347}
Zhao,~Z.; He,~B.; Tang,~B.~Z. Aggregation-Induced Emission of Siloles.
  \emph{Chem. Sci.} \textbf{2015}, \emph{6}, 5347--5365\relax
\mciteBstWouldAddEndPuncttrue
\mciteSetBstMidEndSepPunct{\mcitedefaultmidpunct}
{\mcitedefaultendpunct}{\mcitedefaultseppunct}\relax
\EndOfBibitem
\bibitem[Zhang \latin{et~al.}(2014)Zhang, Jiang, Niu, Wang, Peng, and
  Shuai]{Zhang2014_9094}
Zhang,~T.; Jiang,~Y.; Niu,~Y.; Wang,~D.; Peng,~Q.; Shuai,~Z. Aggregation
  Effects on the Optical Emission of 1,1,2,3,4,5-Hexaphenylsilole (HPS): a
  QM/MM Study. \emph{J. Phys. Chem. A} \textbf{2014}, \emph{118},
  9094--9104\relax
\mciteBstWouldAddEndPuncttrue
\mciteSetBstMidEndSepPunct{\mcitedefaultmidpunct}
{\mcitedefaultendpunct}{\mcitedefaultseppunct}\relax
\EndOfBibitem
\bibitem[Cornil \latin{et~al.}(2001)Cornil, Beljonne, Calbert, and
  Br{\'e}das]{Cornil2001_1053}
Cornil,~J.; Beljonne,~D.; Calbert,~J.-P.; Br{\'e}das,~J.-L. Interchain
  Interactions in Organic $\pi$-Conjugated Materials: Impact on Electronic
  Structure, Optical Response, and Charge Transport. \emph{Adv. Mater.}
  \textbf{2001}, \emph{13}, 1053--1067\relax
\mciteBstWouldAddEndPuncttrue
\mciteSetBstMidEndSepPunct{\mcitedefaultmidpunct}
{\mcitedefaultendpunct}{\mcitedefaultseppunct}\relax
\EndOfBibitem
\bibitem[R{\"o}sch \latin{et~al.}(2006)R{\"o}sch, Yao, Wortmann, and
  W{\"u}rthner]{Rosch2006_7184}
R{\"o}sch,~U.; Yao,~S.; Wortmann,~R.; W{\"u}rthner,~F. Fluorescent H-aggregates
  of Merocyanine Dyes. \emph{Angew. Chem.} \textbf{2006}, \emph{118},
  7184--7188\relax
\mciteBstWouldAddEndPuncttrue
\mciteSetBstMidEndSepPunct{\mcitedefaultmidpunct}
{\mcitedefaultendpunct}{\mcitedefaultseppunct}\relax
\EndOfBibitem
\bibitem[Yoon \latin{et~al.}(2010)Yoon, Chung, Gierschner, Kim, Choi, Kim, and
  Park]{Yoon2010_13675}
Yoon,~S.-J.; Chung,~J.~W.; Gierschner,~J.; Kim,~K.~S.; Choi,~M.-G.; Kim,~D.;
  Park,~S.~Y. Multistimuli Two-Color Luminescence Switching via Different
  Slip-Stacking of Highly Fluorescent Molecular Sheets. \emph{J. Am. Chem.
  Soc.} \textbf{2010}, \emph{132}, 13675--13683\relax
\mciteBstWouldAddEndPuncttrue
\mciteSetBstMidEndSepPunct{\mcitedefaultmidpunct}
{\mcitedefaultendpunct}{\mcitedefaultseppunct}\relax
\EndOfBibitem
\bibitem[Yao and Ashiba(2011)Yao, and Ashiba]{Yao2011_834}
Yao,~H.; Ashiba,~K. Highly Fluorescent Organic Nanoparticles of Thiacyanine
  Dye: A Synergetic Effect of Intermolecular H-aggregation and Restricted
  Intramolecular Rotation. \emph{RSC Adv.} \textbf{2011}, \emph{1},
  834--838\relax
\mciteBstWouldAddEndPuncttrue
\mciteSetBstMidEndSepPunct{\mcitedefaultmidpunct}
{\mcitedefaultendpunct}{\mcitedefaultseppunct}\relax
\EndOfBibitem
\bibitem[Wang \latin{et~al.}(2014)Wang, Shen, Yang, Zhang, Xu, Zhu, Wu, Tian,
  and Zhou]{Wang2014_8723}
Wang,~L.; Shen,~Y.; Yang,~M.; Zhang,~X.; Xu,~W.; Zhu,~Q.; Wu,~J.; Tian,~Y.;
  Zhou,~H. Novel Highly Emissive H-aggregates with Aggregate Fluorescence
  Change in a Phenylbenzoxazole-Based System. \emph{Chem. Commun.}
  \textbf{2014}, \emph{50}, 8723--8726\relax
\mciteBstWouldAddEndPuncttrue
\mciteSetBstMidEndSepPunct{\mcitedefaultmidpunct}
{\mcitedefaultendpunct}{\mcitedefaultseppunct}\relax
\EndOfBibitem
\bibitem[Basak \latin{et~al.}(2015)Basak, Nandi, Bhattacharyya, Datta, and
  Banerjee]{Basak2015_30398}
Basak,~S.; Nandi,~N.; Bhattacharyya,~K.; Datta,~A.; Banerjee,~A. Fluorescence
  from an H-aggregated Naphthalenediimide Based Peptide: Photophysical and
  Computational Investigation of This Rare Phenomenon. \emph{Phys. Chem. Chem.
  Phys.} \textbf{2015}, \emph{17}, 30398--30403\relax
\mciteBstWouldAddEndPuncttrue
\mciteSetBstMidEndSepPunct{\mcitedefaultmidpunct}
{\mcitedefaultendpunct}{\mcitedefaultseppunct}\relax
\EndOfBibitem
\bibitem[Lucenti \latin{et~al.}(2017)Lucenti, Forni, Botta, Carlucci, Giannini,
  Marinotto, Previtali, Righetto, and Cariati]{Lucenti2017_1894}
Lucenti,~E.; Forni,~A.; Botta,~C.; Carlucci,~L.; Giannini,~C.; Marinotto,~D.;
  Previtali,~A.; Righetto,~S.; Cariati,~E. H-Aggregates Granting
  Crystallization-Induced Emissive Behavior and Ultralong Phosphorescence from
  a Pure Organic Molecule. \emph{J. Phys. Chem. Lett.} \textbf{2017}, \emph{8},
  1894--1898\relax
\mciteBstWouldAddEndPuncttrue
\mciteSetBstMidEndSepPunct{\mcitedefaultmidpunct}
{\mcitedefaultendpunct}{\mcitedefaultseppunct}\relax
\EndOfBibitem
\bibitem[Qian \latin{et~al.}(2017)Qian, Cousins, Horak, Wakefield, Liptak, and
  Aprahamian]{Qian2017_83}
Qian,~H.; Cousins,~M.~E.; Horak,~E.~H.; Wakefield,~A.; Liptak,~M.~D.;
  Aprahamian,~I. Suppression of Kasha's Rule as a Mechanism for Fluorescent
  Molecular Rotors and Aggregation-Induced Emission. \emph{Nat. Chem.}
  \textbf{2017}, \emph{9}, 83\relax
\mciteBstWouldAddEndPuncttrue
\mciteSetBstMidEndSepPunct{\mcitedefaultmidpunct}
{\mcitedefaultendpunct}{\mcitedefaultseppunct}\relax
\EndOfBibitem
\bibitem[Ryu \latin{et~al.}(2017)Ryu, Okazaki, Pouget, Takafuji, Nagaoka,
  Ihara, and Oda]{Ryu2017_8870}
Ryu,~N.; Okazaki,~Y.; Pouget,~E.; Takafuji,~M.; Nagaoka,~S.; Ihara,~H.; Oda,~R.
  Fluorescence Emission Originated from the H-aggregated Cyanine Dye with
  Chiral Gemini Surfactant Assemblies Having a Narrow Absorption Band and a
  Remarkably Large Stokes Shift. \emph{Chem. Commun.} \textbf{2017}, \emph{53},
  8870--8873\relax
\mciteBstWouldAddEndPuncttrue
\mciteSetBstMidEndSepPunct{\mcitedefaultmidpunct}
{\mcitedefaultendpunct}{\mcitedefaultseppunct}\relax
\EndOfBibitem
\bibitem[Kasha(1950)]{Kasha1950_14}
Kasha,~M. Characterization of Electronic Transitions in Complex Molecules.
  \emph{Disc. Faraday Soc.} \textbf{1950}, \emph{9}, 14--19\relax
\mciteBstWouldAddEndPuncttrue
\mciteSetBstMidEndSepPunct{\mcitedefaultmidpunct}
{\mcitedefaultendpunct}{\mcitedefaultseppunct}\relax
\EndOfBibitem
\bibitem[Sato(2016)]{Sato2015_189}
Sato,~T. Fluorescence via Reverse Intersystem Crossing from Higher Triplet
  States. \emph{J. Comput. Chem. Jpn.} \textbf{2016}, \emph{14}, 189--192\relax
\mciteBstWouldAddEndPuncttrue
\mciteSetBstMidEndSepPunct{\mcitedefaultmidpunct}
{\mcitedefaultendpunct}{\mcitedefaultseppunct}\relax
\EndOfBibitem
\bibitem[Sato \latin{et~al.}(2017)Sato, Hayashi, Haruta, and Pu]{Sato2017_4820}
Sato,~T.; Hayashi,~R.; Haruta,~N.; Pu,~Y.-J. Fluorescence via Reverse
  Intersystem Crossing from Higher Triplet States in a Bisanthracene
  Derivative. \emph{Sci. Rep.} \textbf{2017}, \emph{7}, 4820\relax
\mciteBstWouldAddEndPuncttrue
\mciteSetBstMidEndSepPunct{\mcitedefaultmidpunct}
{\mcitedefaultendpunct}{\mcitedefaultseppunct}\relax
\EndOfBibitem
\bibitem[Pu \latin{et~al.}(2019)Pu, Satake, Koyama, Otomo, Hayashi, Haruta,
  Katagiri, Otsuki, Kim, and Sato]{Pu2019_2541}
Pu,~Y.-J.; Satake,~R.; Koyama,~Y.; Otomo,~T.; Hayashi,~R.; Haruta,~N.;
  Katagiri,~H.; Otsuki,~D.; Kim,~D.; Sato,~T. Absence of Delayed Fluorescence
  and Triplet--Triplet Annihilation in Organic Light Emitting Diodes with
  Spatially Orthogonal Bianthracenes. \emph{J. Mater. Chem. C} \textbf{2019},
  \emph{7}, 2541--2547\relax
\mciteBstWouldAddEndPuncttrue
\mciteSetBstMidEndSepPunct{\mcitedefaultmidpunct}
{\mcitedefaultendpunct}{\mcitedefaultseppunct}\relax
\EndOfBibitem
\bibitem[Endo \latin{et~al.}(2009)Endo, Ogasawara, Takahashi, Yokoyama, Kato,
  and Adachi]{Endo2009_4802}
Endo,~A.; Ogasawara,~M.; Takahashi,~A.; Yokoyama,~D.; Kato,~Y.; Adachi,~C.
  Thermally activated delayed fluorescence from Sn4+--porphyrin complexes and
  thei r application to organic light emitting diodes―A novel mechanism for
  electroluminescence. \emph{Adv. Mater.} \textbf{2009}, \emph{21},
  4802--4806\relax
\mciteBstWouldAddEndPuncttrue
\mciteSetBstMidEndSepPunct{\mcitedefaultmidpunct}
{\mcitedefaultendpunct}{\mcitedefaultseppunct}\relax
\EndOfBibitem
\bibitem[Adachi(2014)]{Adachi2014_060101}
Adachi,~C. Third-Generation Organic Electroluminescence Materials. \emph{Jpn.
  J. Appl. Phys.} \textbf{2014}, \emph{53}, 060101\relax
\mciteBstWouldAddEndPuncttrue
\mciteSetBstMidEndSepPunct{\mcitedefaultmidpunct}
{\mcitedefaultendpunct}{\mcitedefaultseppunct}\relax
\EndOfBibitem
\bibitem[Hu \latin{et~al.}(2014)Hu, Pu, Satoh, Kawata, Katagiri, Sasabe, and
  Kido]{Hu2014_2064}
Hu,~J.-Y.; Pu,~Y.-J.; Satoh,~F.; Kawata,~S.; Katagiri,~H.; Sasabe,~H.; Kido,~J.
  Bisanthracene-Based Donor--Acceptor-type Light-Emitting Dopants: Highly
  Efficient Deep-Blue Emission in Organic Light-Emitting Devices. \emph{Adv.
  Funct. Mater.} \textbf{2014}, \emph{24}, 2064--2071\relax
\mciteBstWouldAddEndPuncttrue
\mciteSetBstMidEndSepPunct{\mcitedefaultmidpunct}
{\mcitedefaultendpunct}{\mcitedefaultseppunct}\relax
\EndOfBibitem
\bibitem[Nishio \latin{et~al.}(2014)Nishio, Higashiguchi, and
  Matsuda]{Nishio2014_686}
Nishio,~S.; Higashiguchi,~K.; Matsuda,~K. The Effect of Cyano Substitution on
  the Fluorescence Behavior of 1,2-Bis(pyridylphenyl)ethene. \emph{Asian J.
  Org. Chem.} \textbf{2014}, \emph{3}, 686--690\relax
\mciteBstWouldAddEndPuncttrue
\mciteSetBstMidEndSepPunct{\mcitedefaultmidpunct}
{\mcitedefaultendpunct}{\mcitedefaultseppunct}\relax
\EndOfBibitem
\bibitem[Shizu \latin{et~al.}(2013)Shizu, Sato, and Tanaka]{Shizu2013_215}
Shizu,~K.; Sato,~T.; Tanaka,~K. Inverse Relationship of Reorganization Energy
  to The Number of $\pi$ Electrons from Perspective of Vibronic Coupling
  Density. \emph{J. Comput. Chem. Jpn.} \textbf{2013}, \emph{12},
  215--221\relax
\mciteBstWouldAddEndPuncttrue
\mciteSetBstMidEndSepPunct{\mcitedefaultmidpunct}
{\mcitedefaultendpunct}{\mcitedefaultseppunct}\relax
\EndOfBibitem
\bibitem[Sato \latin{et~al.}(2008)Sato, Tokunaga, and Tanaka]{Sato2008_758}
Sato,~T.; Tokunaga,~K.; Tanaka,~K. Vibronic Coupling in Naphthalene Anion:
  Vibronic Coupling Density Analysis for Totally Symmetric Vibrational Modes.
  \emph{J. Phys. Chem. A} \textbf{2008}, \emph{112}, 758--767\relax
\mciteBstWouldAddEndPuncttrue
\mciteSetBstMidEndSepPunct{\mcitedefaultmidpunct}
{\mcitedefaultendpunct}{\mcitedefaultseppunct}\relax
\EndOfBibitem
\bibitem[Sato \latin{et~al.}(2009)Sato, Tokunaga, Iwahara, Shizu, and
  Tanaka]{Sato2009_99}
Sato,~T.; Tokunaga,~K.; Iwahara,~N.; Shizu,~K.; Tanaka,~K. Vibronic Coupling
  Constant and Vibronic Coupling Density. In \emph{The Jahn-Teller Effect:
  Fundamentals and Implications for Physics and Chemistry}; K{\"o}ppel,~H.,
  Yarkony,~D.~R., Barentzen,~H., Eds.; Springer-Verlag: Berlin and Hidelberg,
  2009; pp 99--129\relax
\mciteBstWouldAddEndPuncttrue
\mciteSetBstMidEndSepPunct{\mcitedefaultmidpunct}
{\mcitedefaultendpunct}{\mcitedefaultseppunct}\relax
\EndOfBibitem
\bibitem[Sato \latin{et~al.}(2013)Sato, Uejima, Iwahara, Haruta, Shizu, and
  Tanaka]{Sato2013_012010}
Sato,~T.; Uejima,~M.; Iwahara,~N.; Haruta,~N.; Shizu,~K.; Tanaka,~K. Vibronic
  Coupling Density and Related Concepts. \emph{J. Phys.: Conf. Ser.}
  \textbf{2013}, \emph{428}, 012010\relax
\mciteBstWouldAddEndPuncttrue
\mciteSetBstMidEndSepPunct{\mcitedefaultmidpunct}
{\mcitedefaultendpunct}{\mcitedefaultseppunct}\relax
\EndOfBibitem
\bibitem[Fischer(1984)]{Fischer1984}
Fischer,~G. \emph{Vibronic Coupling: The Interaction between the Electronic and
  Nuclear Motions}; Academic Press: London, 1984\relax
\mciteBstWouldAddEndPuncttrue
\mciteSetBstMidEndSepPunct{\mcitedefaultmidpunct}
{\mcitedefaultendpunct}{\mcitedefaultseppunct}\relax
\EndOfBibitem
\bibitem[Azumi and Matsuzaki(1977)Azumi, and Matsuzaki]{Azumi1977_315}
Azumi,~T.; Matsuzaki,~K. What Does the Term "Vibronic Coupling" Mean?
  \emph{Photochem. Photobiol.} \textbf{1977}, \emph{25}, 315--326\relax
\mciteBstWouldAddEndPuncttrue
\mciteSetBstMidEndSepPunct{\mcitedefaultmidpunct}
{\mcitedefaultendpunct}{\mcitedefaultseppunct}\relax
\EndOfBibitem
\bibitem[Uejima \latin{et~al.}(2014)Uejima, Sato, Yokoyama, Tanaka, and
  Park]{Uejima2014_14244}
Uejima,~M.; Sato,~T.; Yokoyama,~D.; Tanaka,~K.; Park,~J.-W. Quantum Yield in
  Blue-Emitting Anthracene Derivatives: Vibronic Coupling Density and
  Transition Dipole Moment Density. \emph{Phys. Chem. Chem. Phys.}
  \textbf{2014}, \emph{16}, 14244--14256\relax
\mciteBstWouldAddEndPuncttrue
\mciteSetBstMidEndSepPunct{\mcitedefaultmidpunct}
{\mcitedefaultendpunct}{\mcitedefaultseppunct}\relax
\EndOfBibitem
\bibitem[Hutchisson(1930)]{Hutchisson1930_410}
Hutchisson,~E. Band spectra intensities for symmetrical diatomic molecules.
  \emph{Phys. Rev.} \textbf{1930}, \emph{36}, 410\relax
\mciteBstWouldAddEndPuncttrue
\mciteSetBstMidEndSepPunct{\mcitedefaultmidpunct}
{\mcitedefaultendpunct}{\mcitedefaultseppunct}\relax
\EndOfBibitem
\bibitem[Svensson \latin{et~al.}(1996)Svensson, Humbel, Froese, Matsubara,
  Sieber, and Morokuma]{Svensson1996_19357}
Svensson,~M.; Humbel,~S.; Froese,~R.~D.; Matsubara,~T.; Sieber,~S.;
  Morokuma,~K. ONIOM: a multilayered integrated MO+ MM method for geometry
  optimizations and single point energy predictions. A test for Diels- Alder
  reactions and Pt (P (t-Bu) 3) 2+ H2 oxidative addition. \emph{J. Phys. Chem.}
  \textbf{1996}, \emph{100}, 19357--19363\relax
\mciteBstWouldAddEndPuncttrue
\mciteSetBstMidEndSepPunct{\mcitedefaultmidpunct}
{\mcitedefaultendpunct}{\mcitedefaultseppunct}\relax
\EndOfBibitem
\bibitem[Chung \latin{et~al.}(2015)Chung, Sameera, Ramozzi, Page, Hatanaka,
  Petrova, Harris, Li, Ke, Liu, Li, Ding, and Morokuma]{Chung2015_5678}
Chung,~L.~W.; Sameera,~W.; Ramozzi,~R.; Page,~A.~J.; Hatanaka,~M.;
  Petrova,~G.~P.; Harris,~T.~V.; Li,~X.; Ke,~Z.; Liu,~F.; Li,~H.-B.; Ding,~L.;
  Morokuma,~K. The ONIOM Method and its Applications. \emph{Chem. Rev.}
  \textbf{2015}, \emph{115}, 5678--5796\relax
\mciteBstWouldAddEndPuncttrue
\mciteSetBstMidEndSepPunct{\mcitedefaultmidpunct}
{\mcitedefaultendpunct}{\mcitedefaultseppunct}\relax
\EndOfBibitem
\bibitem[Tomasi \latin{et~al.}(2005)Tomasi, Mennucci, and
  Cammi]{Tomasi2005_2999}
Tomasi,~J.; Mennucci,~B.; Cammi,~R. Quantum mechanical continuum solvation
  models. \emph{Chem. Rev.} \textbf{2005}, \emph{105}, 2999--3094\relax
\mciteBstWouldAddEndPuncttrue
\mciteSetBstMidEndSepPunct{\mcitedefaultmidpunct}
{\mcitedefaultendpunct}{\mcitedefaultseppunct}\relax
\EndOfBibitem
\bibitem[Frisch \latin{et~al.}(2013)Frisch, Trucks, Schlegel, Scuseria, Robb,
  Cheeseman, Scalmani, Barone, Mennucci, Petersson, Nakatsuji, Caricato, Li,
  Hratchian, Izmaylov, Bloino, Zheng, Sonnenberg, Hada, Ehara, Toyota, Fukuda,
  Hasegawa, Ishida, Nakajima, Honda, Kitao, Nakai, Vreven, Montgomery~Jr.,
  Peralta, Ogliaro, Bearpark, J, Brothers, Kudin, Staroverov, Keith, Kobayashi,
  Normand, Raghavachari, Rendell, Burant, Iyengar, Tomasi, Cossi, Rega, Millam,
  Klene, Knox, Cross, Bakken, Adamo, Jaramillo, Gomperts, Stratmann, Yazyev,
  Austin, Cammi, Pomelli, Ochterski, Martin, Morokuma, Zakrzewski, Voth,
  Salvador, Dannenberg, Dapprich, Daniels, Farkas, Foresman, Ortiz, Cioslowski,
  and Fox]{Frisch2013D}
Frisch,~M.~J.; Trucks,~G.~W.; Schlegel,~H.~B.; Scuseria,~G.~E.; Robb,~M.~A.;
  Cheeseman,~J.~R.; Scalmani,~G.; Barone,~V.; Mennucci,~B.; Petersson,~G.~A.;
  Nakatsuji,~H.; Caricato,~M.; Li,~X.; Hratchian,~H.~P.; Izmaylov,~A.~F.;
  Bloino,~J.; Zheng,~G.; Sonnenberg,~J.~L.; Hada,~M.; Ehara,~M.; Toyota,~K.;
  Fukuda,~R.; Hasegawa,~J.; Ishida,~M.; Nakajima,~T.; Honda,~Y.; Kitao,~O.;
  Nakai,~H.; Vreven,~T.; Montgomery~Jr.,~J.~A.; Peralta,~J.~E.; Ogliaro,~F.;
  Bearpark,~B.; J,~H.~J.; Brothers,~E.; Kudin,~K.~N.; Staroverov,~V.~N.;
  Keith,~T.; Kobayashi,~R.; Normand,~J.; Raghavachari,~K.; Rendell,~A.;
  Burant,~J.~C.; Iyengar,~S.~S.; Tomasi,~J.; Cossi,~M.; Rega,~N.;
  Millam,~J.~M.; Klene,~M.; Knox,~J.~E.; Cross,~J.~B.; Bakken,~V.; Adamo,~C.;
  Jaramillo,~J.; Gomperts,~R.; Stratmann,~R.~E.; Yazyev,~O.; Austin,~A.~J.;
  Cammi,~R.; Pomelli,~C.; Ochterski,~J.~W.; Martin,~R.~L.; Morokuma,~K.;
  Zakrzewski,~V.~G.; Voth,~G.~A.; Salvador,~P.; Dannenberg,~J.~J.;
  Dapprich,~S.; Daniels,~A.~D.; Farkas,~O.; Foresman,~J.~B.; Ortiz,~J.~V.;
  Cioslowski,~J.; Fox,~D.~J. Gaussian 09, Revision D. 01, Gaussian, Inc.:
  Wallingford, CT, 2013\relax
\mciteBstWouldAddEndPuncttrue
\mciteSetBstMidEndSepPunct{\mcitedefaultmidpunct}
{\mcitedefaultendpunct}{\mcitedefaultseppunct}\relax
\EndOfBibitem
\bibitem[Frisch \latin{et~al.}(2013)Frisch, Trucks, Schlegel, Scuseria, Robb,
  Cheeseman, Scalmani, Barone, Mennucci, Petersson, Nakatsuji, Caricato, Li,
  Hratchian, Izmaylov, Bloino, Zheng, Sonnenberg, Hada, Ehara, Toyota, Fukuda,
  Hasegawa, Ishida, Nakajima, Honda, Kitao, Nakai, Vreven, Montgomery~Jr.,
  Peralta, Ogliaro, Bearpark, J, Brothers, Kudin, Staroverov, Keith, Kobayashi,
  Normand, Raghavachari, Rendell, Burant, Iyengar, Tomasi, Cossi, Rega, Millam,
  Klene, Knox, Cross, Bakken, Adamo, Jaramillo, Gomperts, Stratmann, Yazyev,
  Austin, Cammi, Pomelli, Ochterski, Martin, Morokuma, Zakrzewski, Voth,
  Salvador, Dannenberg, Dapprich, Daniels, Farkas, Foresman, Ortiz, Cioslowski,
  and Fox]{Frisch2013E}
Frisch,~M.~J.; Trucks,~G.~W.; Schlegel,~H.~B.; Scuseria,~G.~E.; Robb,~M.~A.;
  Cheeseman,~J.~R.; Scalmani,~G.; Barone,~V.; Mennucci,~B.; Petersson,~G.~A.;
  Nakatsuji,~H.; Caricato,~M.; Li,~X.; Hratchian,~H.~P.; Izmaylov,~A.~F.;
  Bloino,~J.; Zheng,~G.; Sonnenberg,~J.~L.; Hada,~M.; Ehara,~M.; Toyota,~K.;
  Fukuda,~R.; Hasegawa,~J.; Ishida,~M.; Nakajima,~T.; Honda,~Y.; Kitao,~O.;
  Nakai,~H.; Vreven,~T.; Montgomery~Jr.,~J.~A.; Peralta,~J.~E.; Ogliaro,~F.;
  Bearpark,~B.; J,~H.~J.; Brothers,~E.; Kudin,~K.~N.; Staroverov,~V.~N.;
  Keith,~T.; Kobayashi,~R.; Normand,~J.; Raghavachari,~K.; Rendell,~A.;
  Burant,~J.~C.; Iyengar,~S.~S.; Tomasi,~J.; Cossi,~M.; Rega,~N.;
  Millam,~J.~M.; Klene,~M.; Knox,~J.~E.; Cross,~J.~B.; Bakken,~V.; Adamo,~C.;
  Jaramillo,~J.; Gomperts,~R.; Stratmann,~R.~E.; Yazyev,~O.; Austin,~A.~J.;
  Cammi,~R.; Pomelli,~C.; Ochterski,~J.~W.; Martin,~R.~L.; Morokuma,~K.;
  Zakrzewski,~V.~G.; Voth,~G.~A.; Salvador,~P.; Dannenberg,~J.~J.;
  Dapprich,~S.; Daniels,~A.~D.; Farkas,~O.; Foresman,~J.~B.; Ortiz,~J.~V.;
  Cioslowski,~J.; Fox,~D.~J. Gaussian 09, Revision E. 01, Gaussian, Inc.:
  Wallingford, CT, 2013\relax
\mciteBstWouldAddEndPuncttrue
\mciteSetBstMidEndSepPunct{\mcitedefaultmidpunct}
{\mcitedefaultendpunct}{\mcitedefaultseppunct}\relax
\EndOfBibitem
\bibitem[Sato \latin{et~al.}(2012)Sato, Iwahara, Haruta, and
  Tanaka]{Sato2012_257}
Sato,~T.; Iwahara,~N.; Haruta,~N.; Tanaka,~K. C$_{60}$ bearing ethylene
  moieties. \emph{Chem. Phys. Lett.} \textbf{2012}, \emph{531}, 257--260\relax
\mciteBstWouldAddEndPuncttrue
\mciteSetBstMidEndSepPunct{\mcitedefaultmidpunct}
{\mcitedefaultendpunct}{\mcitedefaultseppunct}\relax
\EndOfBibitem
\bibitem[Yeh \latin{et~al.}(2004)Yeh, Wu, Wen, Dai, Wang, and
  Chen]{Yeh2004_6455}
Yeh,~H.-C.; Wu,~W.-C.; Wen,~Y.-S.; Dai,~D.-C.; Wang,~J.-K.; Chen,~C.-T.
  Derivative of $\alpha$, $\beta$-Dicyanostilbene: Convenient Precursor for the
  Synthesis of Diphenylmaleimide Compounds, E-Z Isomerization, Crystal
  Structure, and Solid-State Fluorescence. \emph{J. Org. Chem.} \textbf{2004},
  \emph{69}, 6455--6462\relax
\mciteBstWouldAddEndPuncttrue
\mciteSetBstMidEndSepPunct{\mcitedefaultmidpunct}
{\mcitedefaultendpunct}{\mcitedefaultseppunct}\relax
\EndOfBibitem
\bibitem[Li \latin{et~al.}(2007)Li, Li, Zhang, Xie, Xie, Xu, Li, Shen, Ye,
  Hanif, Ma, and Ma]{Li2007_231}
Li,~Y.; Li,~F.; Zhang,~H.; Xie,~Z.; Xie,~W.; Xu,~H.; Li,~B.; Shen,~F.; Ye,~L.;
  Hanif,~M.; Ma,~D.; Ma,~Y. Tight Intermolecular Packing through Supramolecular
  Interactions in Crystals of Cyano Substituted Oligo(para-phenylene Vinylene):
  a Key Factor for Aggregation-Induced Emission. \emph{Chem. Commun.}
  \textbf{2007}, 231--233\relax
\mciteBstWouldAddEndPuncttrue
\mciteSetBstMidEndSepPunct{\mcitedefaultmidpunct}
{\mcitedefaultendpunct}{\mcitedefaultseppunct}\relax
\EndOfBibitem
\end{mcitethebibliography}


\providecommand{\latin}[1]{#1}
\providecommand*\mcitethebibliography{\thebibliography}
\csname @ifundefined\endcsname{endmcitethebibliography}
  {\let\endmcitethebibliography\endthebibliography}{}
\begin{mcitethebibliography}{1}
\providecommand*\natexlab[1]{#1}
\providecommand*\mciteSetBstSublistMode[1]{}
\providecommand*\mciteSetBstMaxWidthForm[2]{}
\providecommand*\mciteBstWouldAddEndPuncttrue
  {\def\EndOfBibitem{\unskip.}}
\providecommand*\mciteBstWouldAddEndPunctfalse
  {\let\EndOfBibitem\relax}
\providecommand*\mciteSetBstMidEndSepPunct[3]{}
\providecommand*\mciteSetBstSublistLabelBeginEnd[3]{}
\providecommand*\EndOfBibitem{}
\mciteSetBstSublistMode{f}
\mciteSetBstMaxWidthForm{subitem}{(\alph{mcitesubitemcount})}
\mciteSetBstSublistLabelBeginEnd
  {\mcitemaxwidthsubitemform\space}
  {\relax}
  {\relax}

\bibitem[Atkins \latin{et~al.}(2010)Atkins, Overton, Rourke, Weller, and
  Armstrong]{Atkins2010}
Atkins,~P.; Overton,~T.; Rourke,~J.; Weller,~J.; Armstrong,~F. \emph{Shriver
  and Atkins' Inorganic Chemistry}, 5th ed.; Oxford University Press, USA,
  2010\relax
\mciteBstWouldAddEndPuncttrue
\mciteSetBstMidEndSepPunct{\mcitedefaultmidpunct}
{\mcitedefaultendpunct}{\mcitedefaultseppunct}\relax
\EndOfBibitem
\end{mcitethebibliography}
\bibliographystyle{achemso}

\end{document}


\maketitle
\footnotetext{
 \it 
 $^{a}$~
 Fukui Institute for Fundamental Chemistry, 
 Kyoto University,
 Sakyo-ku,
 Kyoto 606-8103,
 Japan;
 }
\footnotetext{
 \it 
 $^{b}$~
 Department of Molecular Engineering,
 Graduate School of Engineering,
 Kyoto University,
 Nishikyo-ku,
 Kyoto 615-8510,
 Japan
 }
\footnotetext{
 \it 
 $^{c}$~
 Undergraduate School of Industrial Chemistry,
 Faculty of Engineering,
 Kyoto University,
 Nishikyo-ku,
 Kyoto 615-8510,
 Japan
 }
\footnotetext{
 \it
 $^{d}$~
 Department of Synthetic Chemistry and Biological Chemistry, 
 Kyoto University, 
 Nishikyo-ku, 
 Kyoto 615-8510, 
 Japan
 }
\footnotetext{
 \it 
 $^{e}$~
 Unit of Elements Strategy Initiative for Catalysts \& Batteries, 
 Kyoto University,
 Nishikyo-ku,
 Kyoto 615-8510,
 Japan 
 }
 \footnotetext{
 \it
 E-mail: tsato@scl.kyoto-u.ac.jp
 }
\clearpage
\tableofcontents
\clearpage

\section{Dimer Models in Solid Phase}

\vfill
\begin{figure}[!h]
 \centering
 \begin{tabular}{c}
 \includegraphics[scale=0.20]{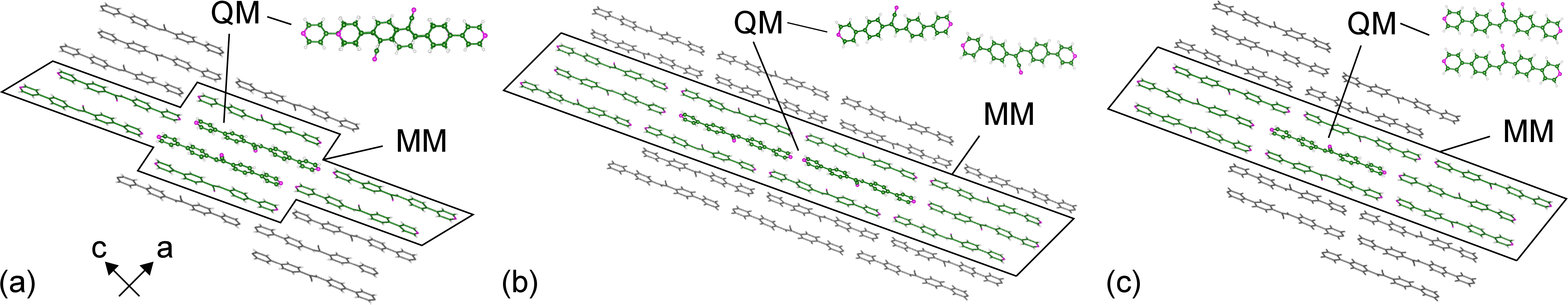} 
 \end{tabular}
 \caption{
  Dimer Model (a) \textbf{1}, (b) \textbf{2}, and (c) \textbf{3} for the CNPPE solid.
  In Dimer Model \textbf{2}, a dimer aligned in the $ca$-plane was selected as the QM region
  and the surrounding 38 molecules were selected as the MM region.
  In Dimer Model \textbf{3}, a dimer aligned in the $b$-direction was selected as the QM region
  and the surrounding 29 molecules were selected as the MM region.
  The QM region was computed at the M06-2X/6-31G(d,p) levels of theory 
  and the MM region was computed using the UFF.
  }
 \label{FIGS1}
\end{figure}

\vfill

\begin{figure}[!h]
 \centering
 \begin{tabular}{c}
 \includegraphics[scale=0.60]{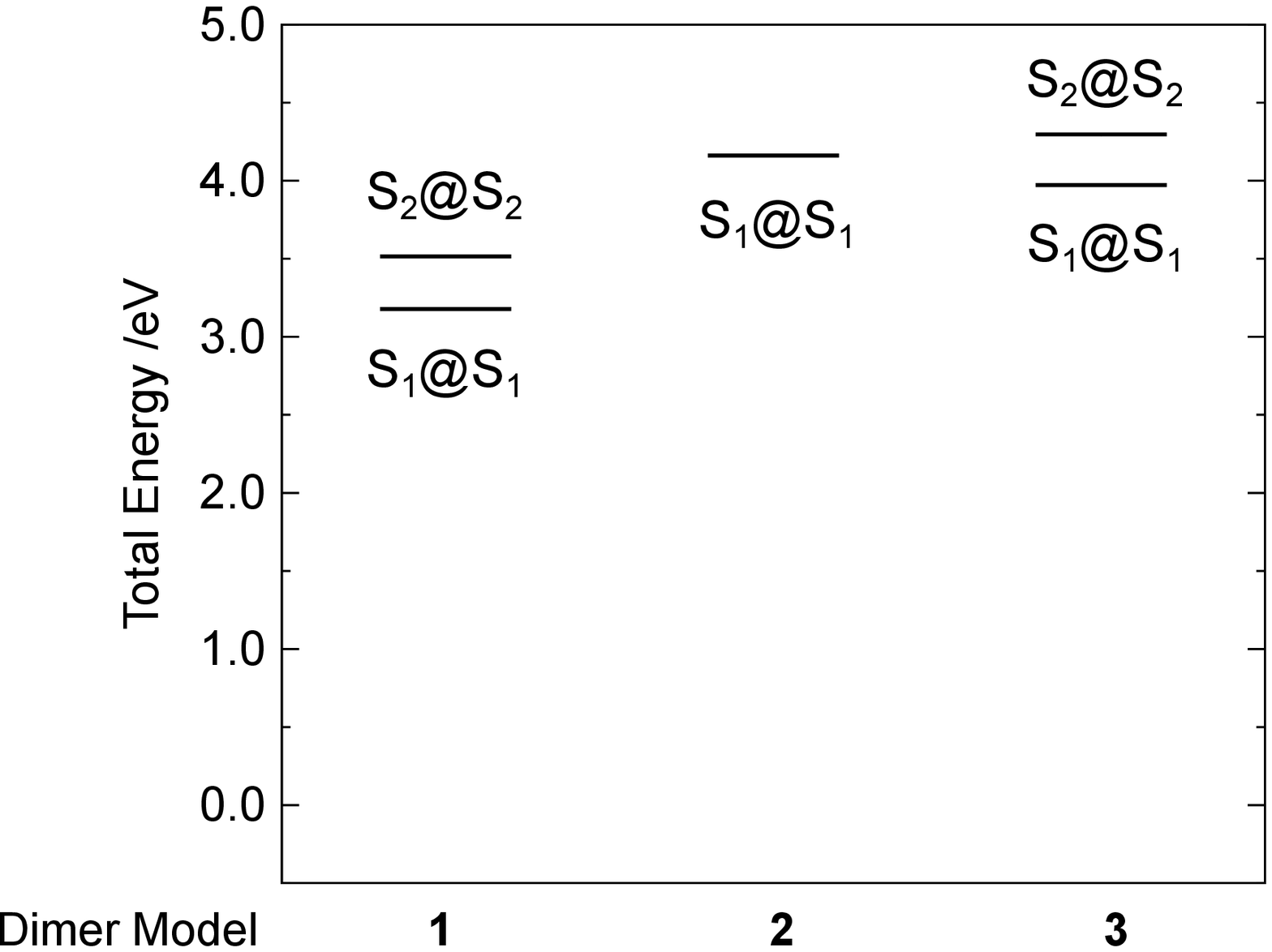} 
 \end{tabular}
 \caption{
 Total energy of the excited states of Dimer Model \textbf{1}, \textbf{2}, and \textbf{3} 
 in the QM region.
 The energy reference is the total energy of S$_0$@S$_0$ of Dimer Model \textbf{1}, -2253.41783105 a.u.
  }
 \label{FIGS2}
\end{figure}

\vfill

\clearpage

\section{Irreducible Representations of Excimers}
\subsection{$C_i$ symmetry}
Table~\ref{TABLES1} presents the character table of $C_i$ point group
\cite{Atkins2010}.
In $C_i$ point group, 
there are cases where an atom does not exist at the inversion center
and where an atom exists at the inversion center.
When an atom does not exist at the inversion center,
the reducible representation of vibrational modes is decomposed as
\begin{equation}
	\Gamma_{{\rm vib}} = \frac{3N-6}{2} A_g + \frac{3N-6}{2} A_u,
\end{equation}
where $N$ denotes the number of atoms.
Half of the vibrational modes belongs to $A_g$ and the other half to $A_u$.
On the other hand, when an atom exists at the inversion center,
the number of vibrational modes is
\begin{equation}
	\Gamma_{{\rm vib}} = \frac{3N-9}{2} A_g + \frac{3N-3}{2} A_u.
\end{equation}
The numbers of vibrational modes belonging to $A_g$ and $A_u$ are different.

\begin{table}[!h]
\centering
\caption{
\label{TABLES1}
Character table of $C_i$ point group.
}
\vspace*{-\intextsep}
\begin{tabular}{ccccc}
\hline\hline
$\bm{C_i}$  & $E$ & $i$ &                  & $h$=2 \\
\hline
$A_g$       &  1  &  1  & $R_x, R_y, R_z$  & $x^2, y^2, z^2, xy, yz, zx$ \\ 
$A_u$       &  1  & -1  & $x, y, z$        &                             \\
\hline\hline
\end{tabular}
\end{table}

\clearpage

\subsection{$C_2$ symmetry}
Table~\ref{TABLES2} presents the character table of $C_2$ point group
\cite{Atkins2010}.
In $C_2$ point group,
there are cases where no atoms exist on a rotational axis and
where atoms exist on a rotational axis.
When atoms do not exist on a rotational axis,
the reducible representation of vibrational modes is decomposed as
\begin{equation}
	\Gamma_{{\rm vib}} = \frac{3N-4}{2} A + \frac{3N-8}{2} B.
\end{equation}
The numbers of vibrational modes belonging to $A$ and $B$ are not the same.
When $N_a$ atoms exist on a rotational axis,
the number of vibrational modes is
\begin{equation}
	\Gamma_{{\rm vib}} = \frac{3N-N_a-4}{2} A + \frac{3N+N_a-8}{2} B.
\end{equation}
The number of vibrational modes belonging to $B$ increases with $N_a$.

\begin{table}[!h]
\centering
\caption{
\label{TABLES2}
Character table of $C_2$ point group.
}
\vspace*{-\intextsep}
\begin{tabular}{ccccc}
\hline\hline
$\bm{C_2}$  & $E$ & $C_2$ &                 & $h$=2 \\
\hline
$A$         &  1  &  1  & $z, R_z$          & $x^2, y^2, z^2, xy$ \\ 
$B$         &  1  & -1  & $x, y, R_x, R_y$  & $yz, zx$            \\
\hline\hline
\end{tabular}
\vspace*{-\intextsep}
\end{table}

\clearpage

\subsection{$C_s$ symmetry}
Table~\ref{TABLES3} presents the character table of $C_s$ point group
\cite{Atkins2010}.
In $C_s$ point group,
there are cases where no atoms exist on a mirror plane
and where atoms exist on a mirror plane.
When atoms do not exist on a mirror plane,
the reducible representation of vibrational modes is decomposed as
\begin{equation}
	\Gamma_{{\rm vib}} = \frac{3N-6}{2} A' + \frac{3N-6}{2} A''.
\end{equation}
Half of the vibrational modes belongs to $A'$ and the other half to $A''$.
When $N_p$ atoms exist on a mirror plane,
the number of vibrational modes is
\begin{equation}
	\Gamma_{{\rm vib}} = \frac{3N+N_p-6}{2} A' + \frac{3N-N_p-6}{2} A''.
\end{equation}
The number of vibrational modes belonging to $A'$ increases with $N_p$.

\begin{table}[!h]
\centering
\caption{
\label{TABLES3}
Character table of $C_s$ point group.
}
\vspace*{-\intextsep}
\begin{tabular}{ccccc}
\hline\hline
$\bm{C_s}$   & $E$ & $\sigma_h$ &                & $h$=2 \\
\hline
$A'$         &  1  &  1         & $x, y, R_z$    & $x^2, y^2, z^2, xy$ \\ 
$A''$        &  1  & -1         & $z, R_x, R_y$  & $yz, zx$            \\
\hline\hline
\end{tabular}
\vspace*{-\intextsep}
\end{table}

\clearpage

\section{
Rate Constant of Internal Conversion for Monomer Model and Dimer Model \textbf{1}
}

~\\

\begin{table}[!h]
\centering
\vspace*{-\intextsep}
\caption{\label{TABLES4}
Dependence of $k_{n \leftarrow m}^{{\rm IC}} (T)$ within a single mode approximation
on the linewidth of the Gaussian function $\sigma$.
$V_{mn,\alpha}=5\times10^{-4}$ a.u.,
$\omega_{\alpha}=1700$ cm$^{-1}$, 
$E_m - E_n=3.0$ eV,
$E_{{\rm min}}=-11500$ cm$^{-1}$, and
$E_{{\rm max}}=11500$ cm$^{-1}$.
$V_{n,\alpha}$ at maximum coupling mode of Monomer model is $8.42\times10^{-4}$ a.u.
and of Dimer Model \textbf{1} is $5.27\times10^{-4}$ a.u.
Ratios of $k_{n \leftarrow m}^{{\rm IC}} (T)$ of Monomer Model to Dimer Model \textbf{1}
are also given.
}
\scalebox{0.8}{
\begin{tabular}{cccccc}
\hline\hline
                      & \multicolumn{2}{c}{$k_{n\leftarrow m}^{{\rm IC}}$ (s$^{-1}$)} \\
\cline{2-3} 
 $\sigma$ (cm$^{-1}$) & Monomer Model & Dimer Model \textbf{1}         & Ratio \\
\hline
300 & $1.195\times 10^{10}$ & $2.998\times 10^7$ & 388\\
400 & $1.335\times 10^{10}$ & $3.985\times 10^7$ & 335\\
500 & $1.569\times 10^{10}$ & $5.651\times 10^7$ & 277\\
600 & $1.832\times 10^{10}$ & $7.552\times 10^7$ & 242\\
700 & $2.092\times 10^{10}$ & $9.525\times 10^7$ & 219\\
\hline\hline
\end{tabular}
}
\end{table}

\clearpage

\section{Electron Density Differences of Dimer Model \textbf{2} and \textbf{3}}

\begin{table}[!h]
	\centering
\vspace*{-\intextsep}
	\caption{
	Excited states of Dimer Model \textbf{2} at the S$_0$ and S$_1$ optimized structures.
	\label{TABLES5}
		}
\scalebox{.7}{
\begin{tabular}{cccccc}
\hline\hline
	  &  State &   \multicolumn{2}{c}{Excitation Energy}  & $f$ & Major Configuration \\
	 \cline{3-4} 
	  &        &   eV  &  nm        &            &          (CI Coefficient)\\
 \hline 
	  @S$_0$ & S$_1$ ($A$)  & 3.8692 & 320.44 &  3.2829 & HO-1 $\rightarrow$ LU (0.398),
                                                              HO $\rightarrow$ LU+1 (0.565) \\
		 & S$_2$ ($A$)  & 3.9294 & 315.53 &  0.0977 & HO-1 $\rightarrow$ LU (0.565),
                                                              HO $\rightarrow$ LU+1 (-0.399) \\
	  @S$_1$ & S$_1$ ($A$)  & 3.1122 & 398.39 &  1.8360 & HO $\rightarrow$ LU (0.700) \\
		 & S$_2$ ($A$)  & 3.9107 & 317.04 &  1.5560 & HO-1 $\rightarrow$ LU+1 (0.791)\\
\hline\hline
\end{tabular}
} 
\end{table}

\vfill

\begin{figure}[!h]
 \centering
 \begin{tabular}{c}
 \includegraphics[scale=0.30]{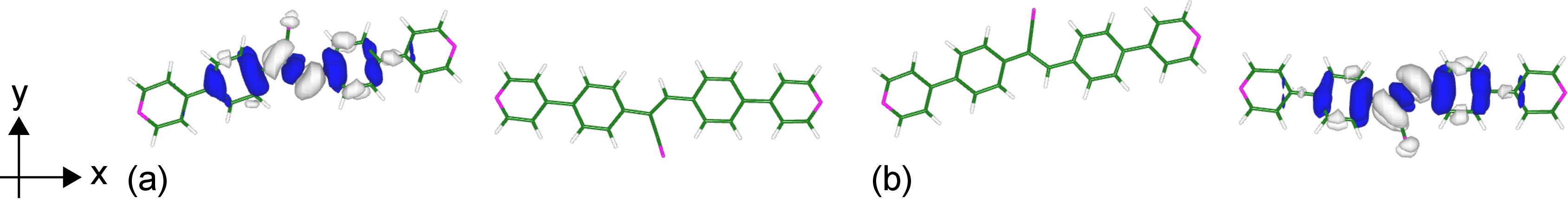} 
 \end{tabular}
 \caption{
	Electron density differences of 
	(a) S$_1$@S$_1$-S$_0$@S$_1$ and (b) S$_2$@S$_1$-S$_0$@S$_1$ 
	for Dimer Model \textbf{2}.
	Isosurface values are $1.0\times10^{-3}$ a.u.
  }
 \label{FIGS3}
\end{figure}

\vfill

\begin{table}[!h]
	\centering
\vspace*{-\intextsep}
	\caption{
	Excited states of Dimer Model \textbf{3} at the S$_0$ and S$_2$ optimized structures.
	\label{TABLES6}
		}
\scalebox{.7}{
\begin{tabular}{cccccc}
\hline\hline
	  &  State &   \multicolumn{2}{c}{Excitation Energy}  & $f$ & Major Configuration \\
	 \cline{3-4} 
	  &        &   eV  &  nm        &            &          (CI Coefficient)\\
 \hline 
          @S$_0$ & S$_{1}$ ($A$) & 3.7862 & 327.46 & 0.1146 & HO-1 $\rightarrow$ LU ( 0.57769),
                                                              HO $\rightarrow$ LU+1 (-0.37639)\\
                 & S$_{2}$ ($A$) & 3.9346 & 315.12 & 2.6588 & HO-1 $\rightarrow$ LU ( 0.38099),
                                                              HO  $\rightarrow$ LU+1 ( 0.57373)\\ 
          @S$_2$ & S$_{1}$ ($A$) & 3.3696 & 367.95 & 0.0005 & HO-1 $\rightarrow$ LU ( 0.49586),
                                                              HO $\rightarrow$ LU+1 (-0.48388)\\
                 & S$_{2}$ ($A$) & 3.5310 & 351.13 & 2.8119 & HO-1 $\rightarrow$ LU ( 0.48848),
                                                              HO   $\rightarrow$ LU+1 ( 0.49548)\\
\hline\hline
\end{tabular}
} 
\end{table}

\vfill

\begin{figure}[!h]
 \centering
 \begin{tabular}{c}
 \includegraphics[scale=0.35]{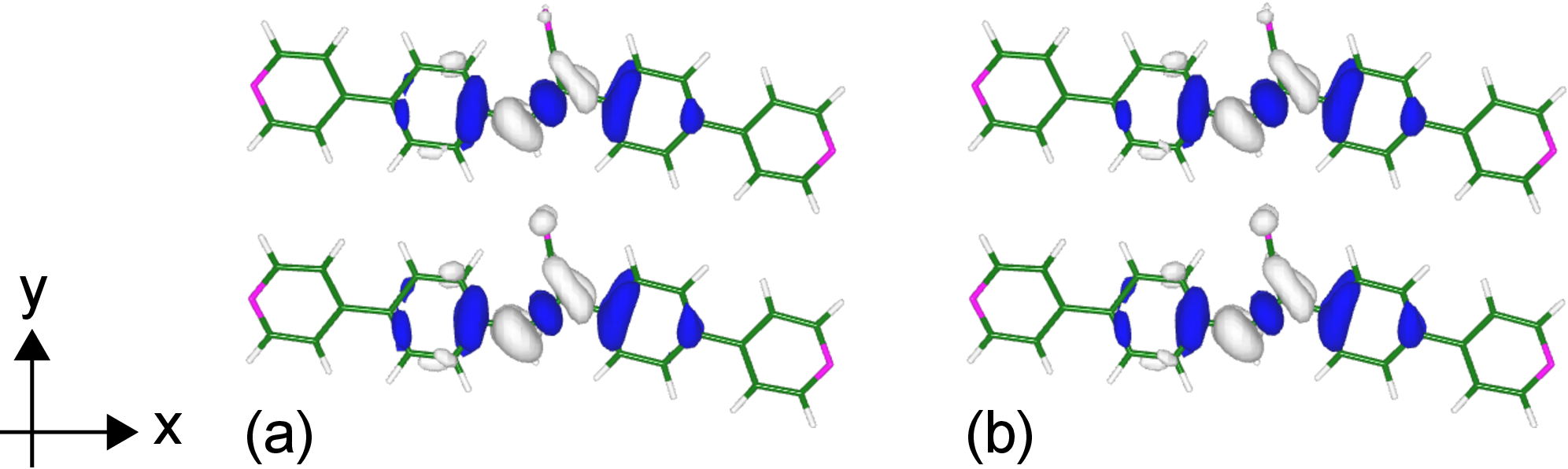} 
 \end{tabular}
 \caption{
	Electron density differences of 
	(a) S$_1$@S$_2$-S$_0$@S$_2$ and (b) S$_2$@S$_2$-S$_0$@S$_2$ 
	for Dimer Model \textbf{3}.
	Isosurface values are $1.0\times10^{-3}$ a.u.
  }
 \label{FIGS4}
\end{figure}

\vfill

\clearpage

\section{Potential Derivatives of Monomer Model and Dimer Model \textbf{1}}

\begin{figure}[!h]
 \centering
 \includegraphics[scale=0.20]{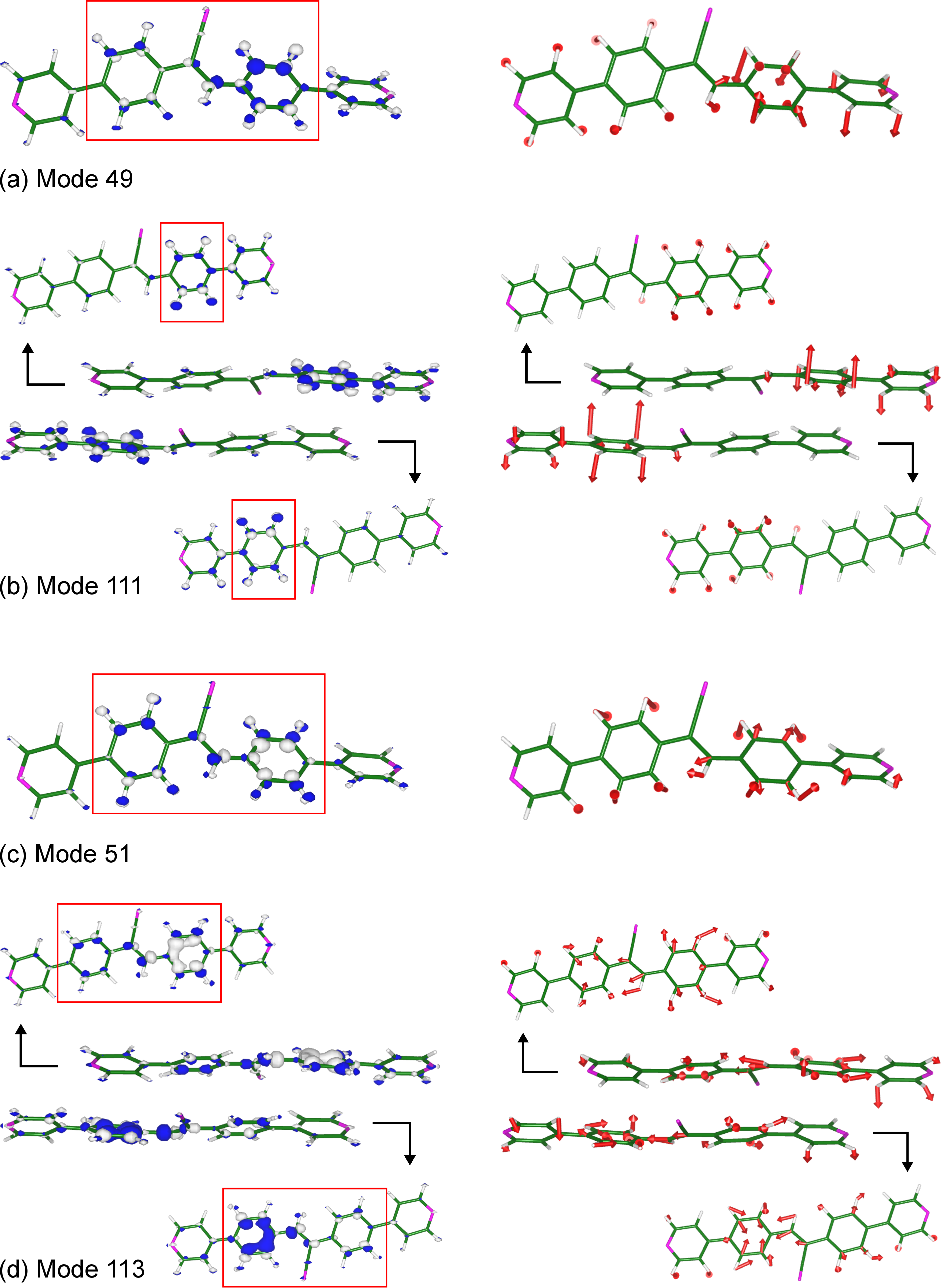} 
 \caption{
 Potential derivatives of 
 (a) mode 49 for Monomer Model and (b) mode 111 for Dimer Model \textbf{1}
 of which the off-diagonal VCC is reduced by the packing effect.
 Potential derivatives of 
 (c) mode 51 for Monomer Model and (d) mode 113 for Dimer Model \textbf{1}
 of which the off-diagonal VCC is not reduced by the packing effect.
 Isosurface value for Monomer Model is $1\times10^{-2}$ a.u.
 and for Dimer Model \textbf{1} is $1/\sqrt{2}\times10^{-2}$ a.u.
  }
 \label{FIGS5}
\end{figure}

\clearpage

\section{Potential Energy Surface of Monomer Model}

\begin{figure}[!h]
 \centering
 \includegraphics[scale=0.5]{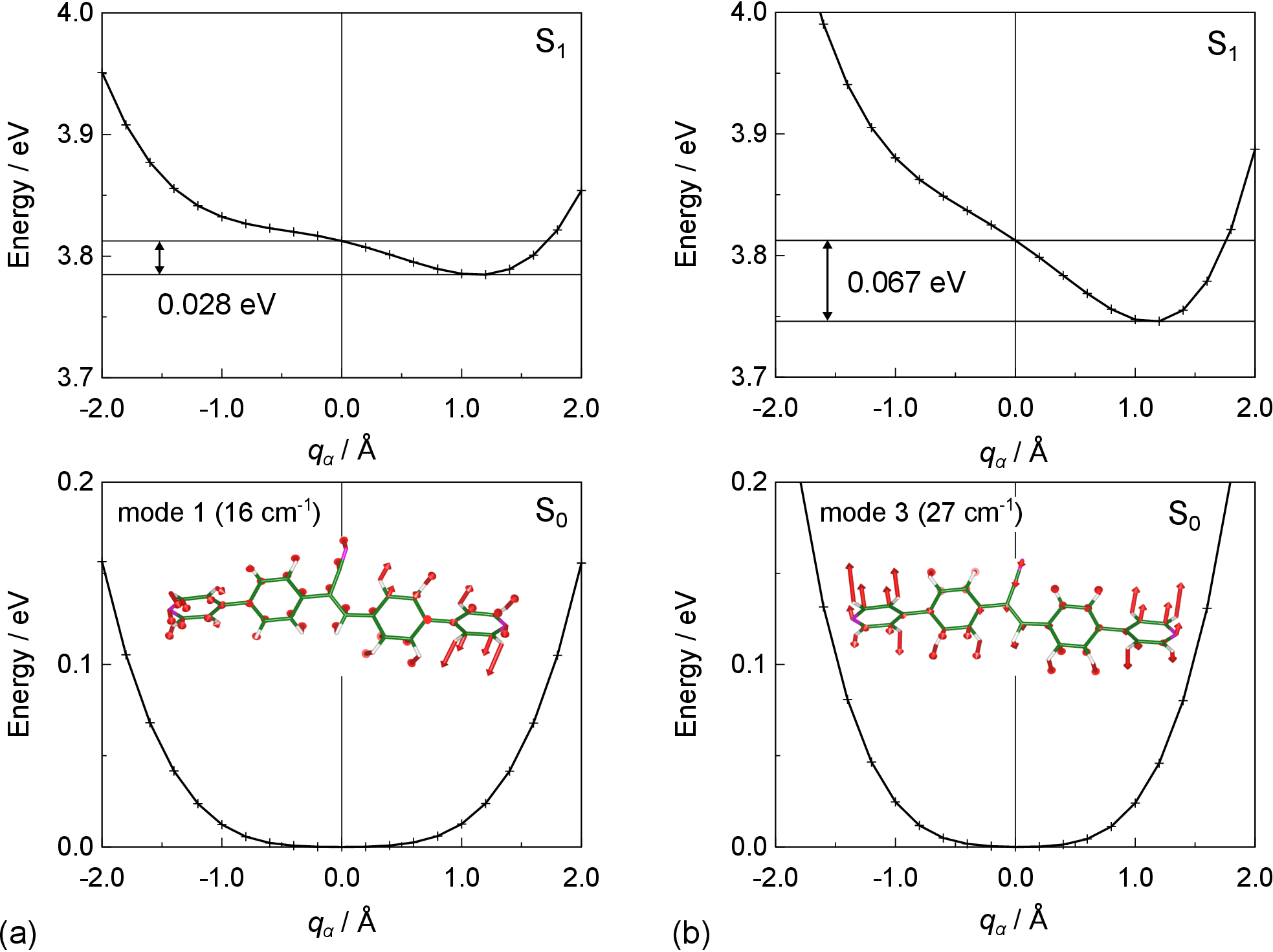} 
 \caption{
  Potential energy surfaces of S$_0$ and S$_1$ of Monomer Model  
  along with (a) mode 1 and (b) mode 3.
  $q_{\alpha}$ denotes the Cartesian displacement from the S$_0$ optimized structure.
  The potential energy surfaces along with these modes were not approximated to harmonic potentials.
  The reorganization energies of modes 1 and 3 including the anharmonicity effect 
  were calculated to be 0.028 and 0.067 eV, respectively.
  }
 \label{FIGS6}
\end{figure}

\clearpage

\section{
Decamer Model in Solid Phase
}

~\\

\vfill

\begin{figure}[!h]
 \centering
 \includegraphics[scale=0.4]{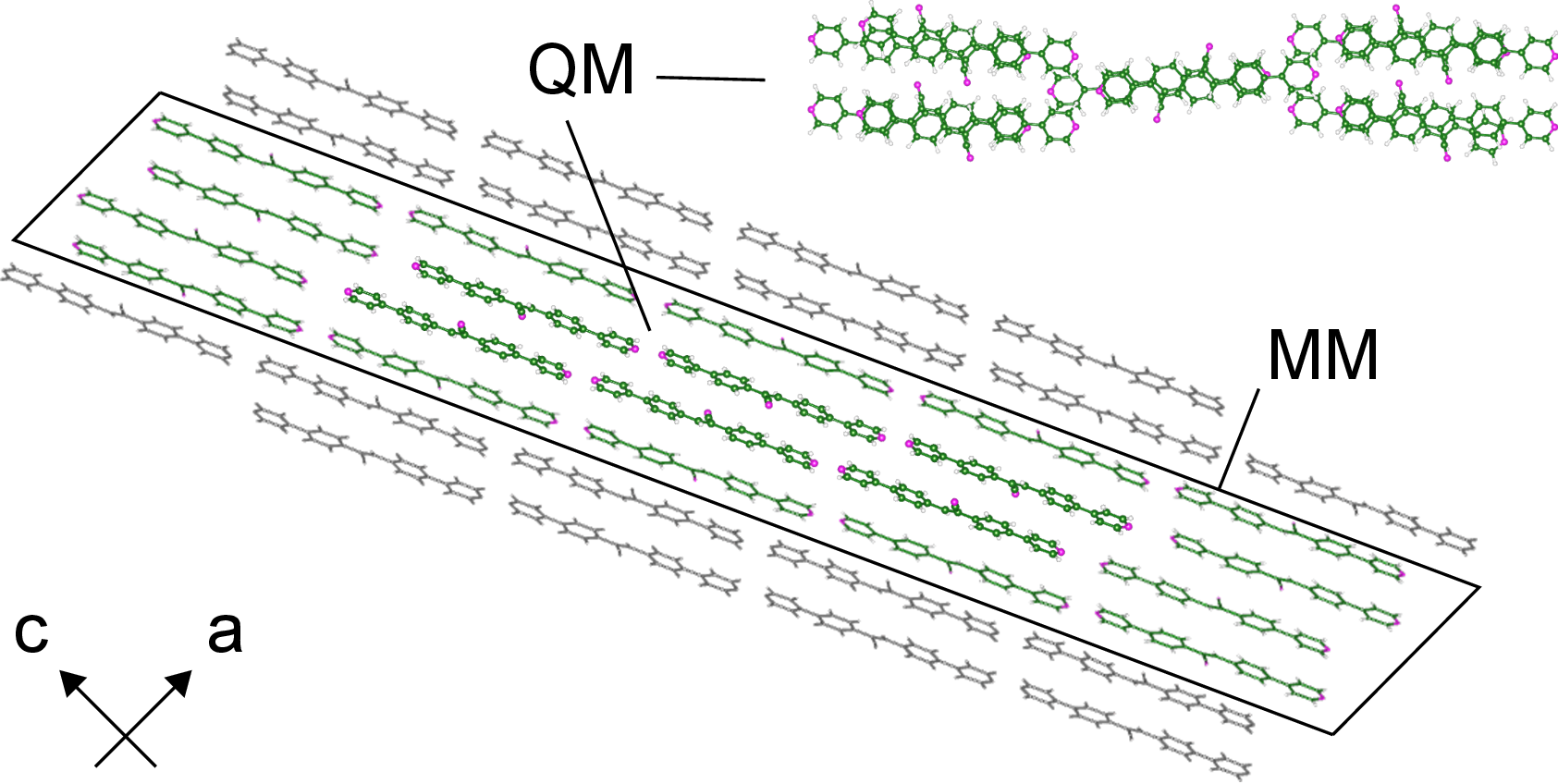} 
 \caption{
 Decamer Model for the CNPPE solid. 
 Centered decamer is selected as the QM region (M06-2X/3-21G) 
 and the surrounding 60 molecules are selected as the MM region (UFF). 
}
 \label{FIGS7}
\end{figure}

\vfill

\begin{figure}[!h]
 \centering
 \includegraphics[scale=0.4]{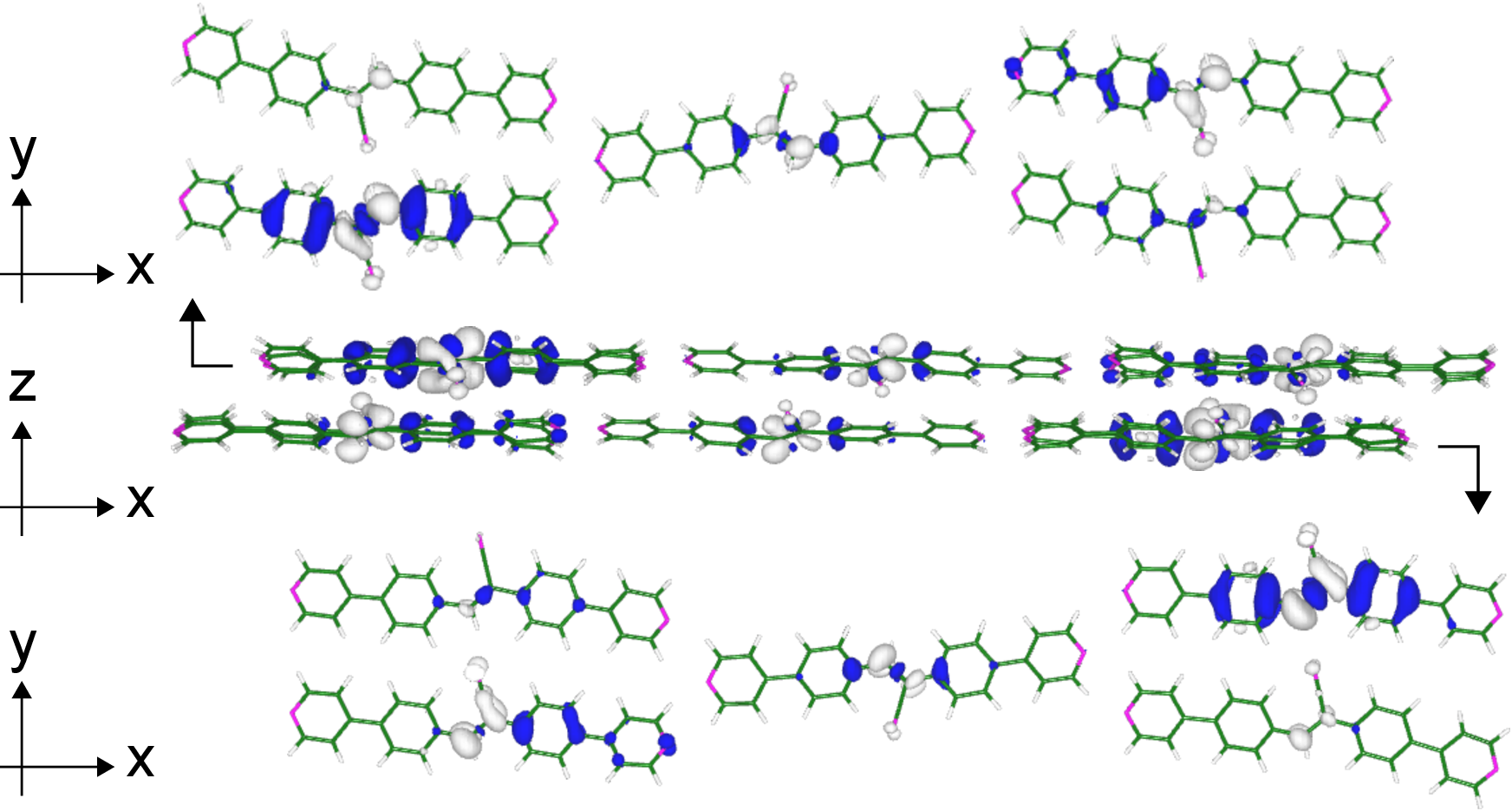} 
 \caption{
 Electron density difference of S$_8$@S$_0$-S$_0$@S$_0$ for Decamer Model. 
 Isosurface value is $3.0\times10^{-4}$ a.u.
  }
 \label{FIGS8}
\end{figure}

\vfill

\clearpage

\begin{table}[!h]
\centering
\caption{
\label{TABLES7}
Excited states of Decamer Model at the S$_0$ optimized structure.
}
\scalebox{.8}{
\begin{tabular}{cccccccc}
\hline\hline
           State   & \multicolumn{2}{c}{Excitation Energy}  & $f$ & Major Configuration \\
         \cline{2-3} 
                   & eV  &  nm        &            &          (CI Coefficient)\\
 \hline 
         S$_1$ ($A_u$)  & 3.7447 & 331.09 &  0.0052 & HO-4 $\rightarrow$ LU (0.2715),
                                                      HO-4 $\rightarrow$ LU+5 (-0.2389) \\
                        &        &        &         & HO-3 $\rightarrow$ LU+1 (-0.2715),
                                                      HO-3 $\rightarrow$ LU+4 (-0.2389) \\
         S$_2$ ($A_g$)  & 3.7447 & 331.09 &  0.0000 & HO-4 $\rightarrow$ LU+1 (0.2717),
                                                      HO-4 $\rightarrow$ LU+4 (0.2392) \\
                        &        &        &         & HO-3 $\rightarrow$ LU   (-0.2717),
                                                      HO-3 $\rightarrow$ LU+5 (0.2392) \\
         S$_3$ ($A_g$)  & 3.7775 & 328.21 &  0.0000 & HO-4 $\rightarrow$ LU+1 (0.2197),
                                                      HO-3 $\rightarrow$ LU   (-0.2199) \\
                        &        &        &         & HO-2 $\rightarrow$ LU+3 (0.2583),
                                                      HO-1 $\rightarrow$ LU+2 (0.2583) \\
         S$_4$ ($A_u$)  & 3.7775 & 328.21 &  0.0618 & HO-4 $\rightarrow$ LU   (-0.2204),
                                                      HO-3 $\rightarrow$ LU+1 (0.2204) \\
                        &        &        &         & HO-2 $\rightarrow$ LU+2 (0.2580),
                                                      HO-1 $\rightarrow$ LU+3 (0.2580) \\
         S$_5$ ($A_u$)  & 3.8860 & 319.06 &  0.7377 & HO-4 $\rightarrow$ LU+2 (-0.2070),
                                                      HO-4 $\rightarrow$ LU+5 (0.2771) \\
                        &        &        &         & HO-3 $\rightarrow$ LU+3 (-0.2070),
                                                      HO-3 $\rightarrow$ LU+4 (0.2771) \\
                        &        &        &         & HO-2 $\rightarrow$ LU+6 (-0.2052),
                                                      HO-1 $\rightarrow$ LU+7 (0.2052) \\
         S$_6$ ($A_g$)  & 3.8871 & 318.96 &  0.0000 & HO-4 $\rightarrow$ LU+3 (-0.2030),
                                                      HO-4 $\rightarrow$ LU+4 (0.2690) \\
                        &        &        &         & HO-3 $\rightarrow$ LU+2 (-0.2030),
                                                      HO-3 $\rightarrow$ LU+5 (0.2690) \\
                        &        &        &         & HO-2 $\rightarrow$ LU+7 (0.2208),
                                                      HO-1 $\rightarrow$ LU+6 (-0.2208) \\
         S$_7$ ($A_g$)  & 3.9259 & 315.81 &  0.0000 & HO-5 $\rightarrow$ LU+9 (-0.2354),
                                                      HO   $\rightarrow$ LU+8 (0.6484) \\
         S$_8$ ($A_u$)  & 3.9942 & 310.41 &  9.0385 & HO-2 $\rightarrow$ LU+6 (-0.3474),
                                                    HO-1 $\rightarrow$ LU+7 (0.3474) \\
                      &        &        &          &HO   $\rightarrow$ LU+9 (-0.2498) \\
        S$_{9}$ ($A_g$) & 4.0290 & 307.73 & 0.0000 & HO-2 $\rightarrow$ LU+7 (-0.3649),
                                                     HO-1 $\rightarrow$ LU+6 ( 0.3649)\\
        S$_{10}$ ($A_u$) & 4.1224 & 300.76 & 1.1521 & HO $\rightarrow$ LU+9 (-0.5282)\\
\hline\hline
\end{tabular}
} 
\end{table}

\clearpage

\section{
Dimer Model \textbf{1} Calculated by $\omega$B97X-D Functional
}

~\\
\vfill

\begin{table}[!h]
\vspace*{-\intextsep}
\centering
\caption{
\label{TABLES8}
Excited states of Dimer Model \textbf{1} at the S$_0$ and S$_2$ optimized structures
calculated by $\omega$B97X-D/6-31G(d,p) level of theory.
$f$ denotes the oscillator strength.
}
\scalebox{.8}{
\begin{tabular}{cccccc}
\hline\hline
         &               & \multicolumn{2}{c}{Excitation Energy} &         & Major Configuration \\
 \cline{3-4} 
         & State         & eV     &  nm                          & $f$     & (CI coefficient)\\
 \hline 
      @S$_0$ & S$_1$(A$_g$)  & 3.8193 & 324.63                  &  0.0000 & HO-1 $\rightarrow$ LU+1 (-0.379)\\
             &               &        &                         &         & HO $\rightarrow$ LU (0.564) \\
             & S$_2$(A$_u$)  & 3.9453 & 314.26                  &  2.3698 & HO-1 $\rightarrow$ LU (-0.406)\\
             &               &        &                         &         & HO $\rightarrow$ LU+1 (0.542) \\
      @S$_2$ & S$_1$(A$_g$)  & 3.3319 & 372.12                  &  0.0000 & HO-1 $\rightarrow$ LU+1 (0.347)\\
             &               &        &                         &         & HO $\rightarrow$ LU (0.592) \\
             & S$_2$(A$_u$)  & 3.4697 & 357.33                  &  2.3091 & HO-1 $\rightarrow$ LU (0.379)\\
             &               &        &                         &         & HO $\rightarrow$ LU+1 (0.570)\\
\hline\hline
\end{tabular}
} 
\vspace*{-\intextsep}
\end{table}

~\\

\vfill

\begin{figure}[!h]
\vspace*{-\intextsep}
\centering
 \includegraphics[scale=0.6]{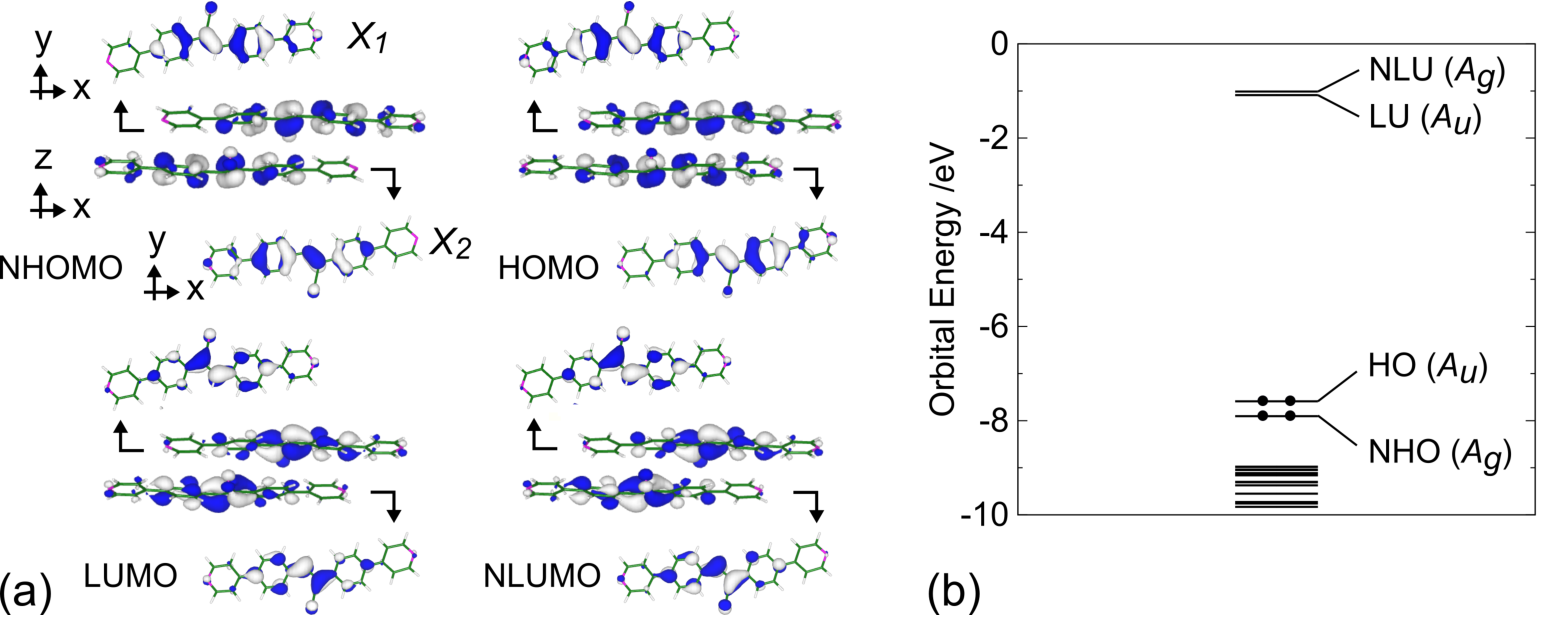} 
 \caption{
(a) Frontier orbitals and 
(b) orbital levels of Dimer Model \textbf{1} at the S$_2$ optimized structure
calculated by $\omega$B97X-D/6-31G(d,p) level of theory.
$X_1$ and $X_2$ are the constituent molecules of Dimer Model \textbf{1}.
Isosurface values of the frontier orbitals are $3.0\times 10^{-2}$ a.u.
}
\label{FIGS9}
\vspace*{-\intextsep}
\end{figure}

\vfill

~\\
\clearpage

\section{Hubbard Model of a Pseudo-Degenerate Electronic System}

\begin{figure}[!h]
 \centering
 \includegraphics[scale=0.45]{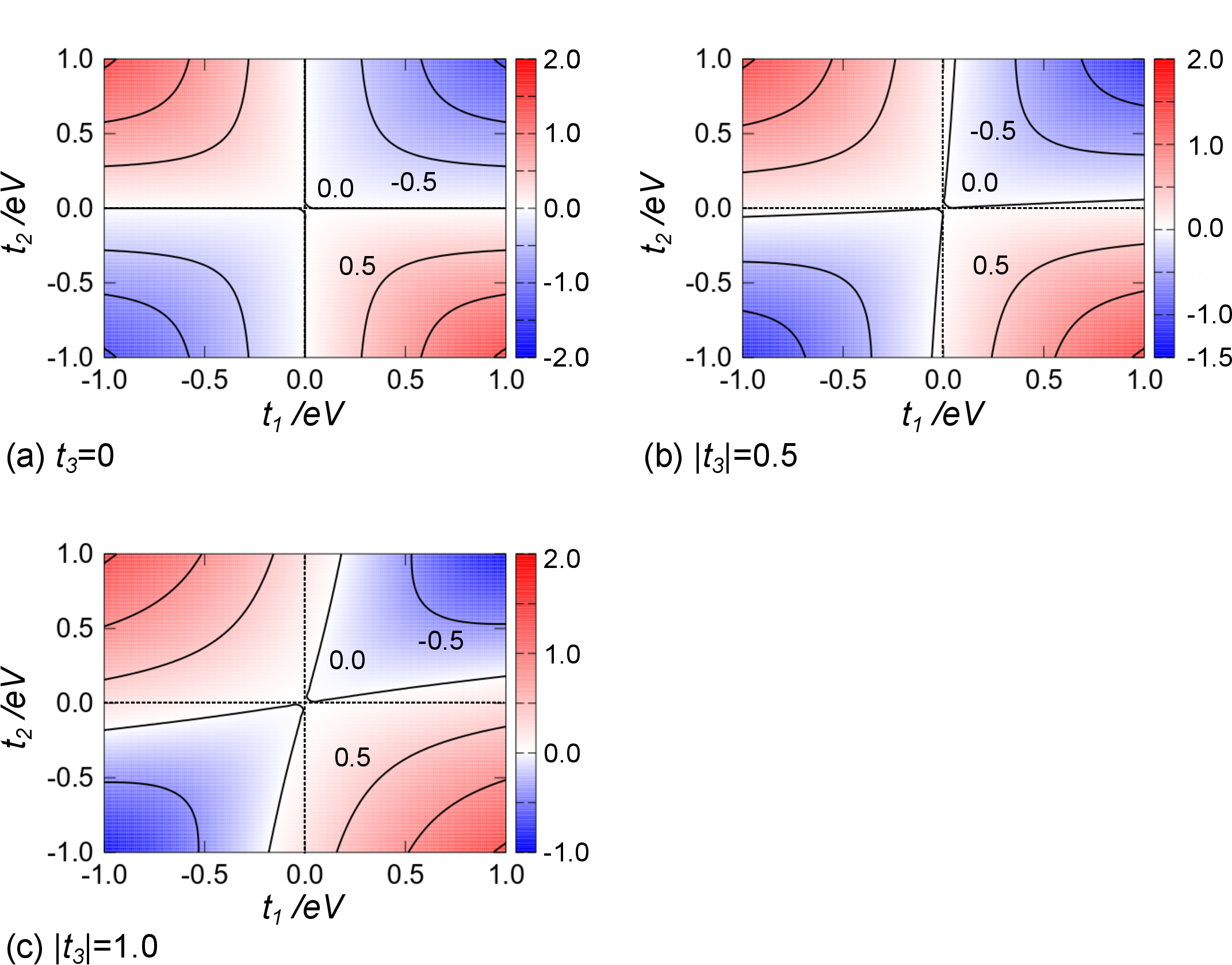} 
 \caption{
  Energy difference between first and second excited states, $E^s_1-E_1^a$,
  calculated by numerically diagonalizing the Hubbard Hamiltonian.
  $\epsilon$, $U_1$, and $U_2$ were set to $3.0$, $1.0$, and $1.0$, respectively.
  }
 \label{FIGS10}
\end{figure}

\clearpage

In this section,
we describe the electron density differences and overlap densities 
in the pseudo-degenerate electronic system using the Hubbard model.
The electron density in the ground state is defined by $\rho_0$.
The orbital overlap densities are defined by
\begin{gather}
p_1 = |\phi_{{\rm HO}} (X_1)|^2, \ \ \
p_2 = |\phi_{{\rm HO}} (X_2)|^2, \ \ \
q_1 = |\phi_{{\rm LU}} (X_1)|^2, \ \ \
q_2 = |\phi_{{\rm LU}} (X_2)|^2, \\
r_1 = \phi_{{\rm HO}} (X_1) \phi_{{\rm LU}} (X_1), \ \ \ 
r_2 = \phi_{{\rm HO}} (X_2) \phi_{{\rm LU}} (X_2), \\
s_1 = \phi_{{\rm HO}} (X_1) \phi_{{\rm HO}} (X_2), \ \ \ 
s_2 = \phi_{{\rm LU}} (X_1) \phi_{{\rm LU}} (X_2), \ \ \ 
s_3 = \phi_{{\rm HO}} (X_1) \phi_{{\rm LU}} (X_2)
= \phi_{{\rm LU}} (X_1) \phi_{{\rm HO}} (X_2),
\end{gather}
where $\phi_{{\rm HO/LU}} (X_1/X_2)$
represent the HOMO/LUMO of $X_1$/$X_2$.
The overlaps between $X_1$ and $X_2$ separated from each other, 
namely $s_1$, $s_2$ and $s_3$, 
are supposed to be 0.
Table~\ref{TABLES9} and \ref{TABLES10} respectively
present the electron density differences and overlap densities 
in the pseudo-degenerate electronic system.

\vfill

\begin{table}[h!]
\vspace*{-\intextsep}
\centering
\caption{
\label{TABLES9}
Electron density differences in the pseudo-degenerate system.
$
\alpha = \frac{1}{2} ( (\rho_0-p_1+q_1)+(\rho_0-p_2+q_2)).
$
For Dimer Model \textbf{1} of the CNPPE solid,
S$_1$ corresponds to $\ket{\Psi_1^s}$, and S$_2$ $\ket{\Psi_1^a}$.
Therefore, 
the electron density difference between S$_1$ and S$_2$ is given by 0.
}
\begin{tabular}{cccccc} 
\hline\hline
             & $\Psi_0$  & $\Psi_1^s$      & $\Psi_2^s$      & $\Psi_1^a$      & $\Psi_2^a$    \\ \hline
$\Psi_0$     & $0$       & $\rho_0-\alpha$ & $\rho_0-\alpha$ & $\rho_0-\alpha$ & $\rho_0-\alpha$ \\ 
$\Psi_1^s$   &           & $0$             & $0$             & $0$             & $0$              \\
$\Psi_2^s$   &           &                 & $0$             & $0$             & $0$              \\
$\Psi_1^a$   &           &                 &                 & $0$             & $0$              \\
$\Psi_2^a$   &           &                 &                 &                 & $0$              \\
\hline\hline
\end{tabular}
\vspace*{-\intextsep}
\end{table}

\vfill

\begin{table}[!h]
\vspace*{-\intextsep}
\centering
\caption{
\label{TABLES10}
Overlap densities in the pseudo-degenerate system.
$
\alpha = \frac{1}{2} ( (\rho_0-p_1+q_1)+(\rho_0-p_2+q_2)),\
\beta_1 = \frac{1}{2} (q_1-p_1-q_2+p_2) \approx 0,\
\beta_2 = \frac{1}{2} (q_2-p_1-q_1+p_2),\
\gamma_1 = \frac{1}{\sqrt{2}} (r_1+r_2),\
\gamma_2 = \frac{1}{\sqrt{2}} (r_1-r_2).
$
For Dimer Model \textbf{1} of the CNPPE solid,
S$_1$ corresponds to $\ket{\Psi_1^s}$, and S$_2$ $\ket{\Psi_1^a}$.
Therefore, 
the overlap density between S$_1$ and S$_2$
is given by $\beta_1 \approx 0$.
}
\begin{tabular}{cccccc} 
\hline\hline
              & $\Psi_0$  & $\Psi_1^s$      & $\Psi_2^s$      & $\Psi_1^a$      & $\Psi_2^a$    \\ \hline
\hline
$\Psi_0$      & $\rho_0$  & $\gamma_1$    & $\frac{t_1+t_2}{U_2}\gamma_1-\frac{t_3}{\epsilon-U_1+2U_2}\alpha$ 
& $\gamma_2$  & $\frac{t_2-t_1}{U_2}\gamma_2-\frac{t_3}{\epsilon-U_1+2U_2}\beta_2$\\ 
$\Psi_1^s$    &           & $\alpha$      & $0$         
& $\beta_1$($\approx$ 0)   & $\frac{t_2-t_1}{U_2}\beta_1-\frac{t_1+t_2}{U_2}\beta_2$  \\
$\Psi_2^s$    &           &               & $\alpha$    
& $\frac{t_1+t_2}{U_2}\beta_1-\frac{t_2-t_1}{U_2}\beta_2$  & $\beta_2$   \\
$\Psi_1^a$    &           &               &             & $\alpha$    & $0$  \\
$\Psi_2^a$    &           &               &             &             & $\alpha$    \\
\hline\hline
\end{tabular}
\vspace*{-\intextsep}
\end{table}

\vfill

\clearpage

\bibliography{refs-ESI} 
\bibliographystyle{achemso} 